\documentclass[a4paper,12pt]{article}

\usepackage{latexsym}
\usepackage{amsmath,amssymb,amsxtra,amscd,amsthm}
\usepackage[mathscr]{eucal}
\usepackage{epsfig}
\usepackage{datetime}
\usepackage{pdfsync}
\usepackage{thumbpdf}
\usepackage{color}

\numberwithin{equation}{section}

\providecommand{\href}[2]{#2}

\def\cD{\mathcal{D}}
\def\cE{\mathcal{E}}
\def\cF{\mathcal{F}}
\def\cL{\mathcal{L}}
\def\cM{\mathcal{M}}
\def\cZ{\mathcal{Z}}
\def\mC{\mathbb{C}}
\def\mF{\mathbb{F}}
\def\mP{\mathbb{P}}
\def\mZ{\mathbb{Z}}
\def\tS{\mathrm{S}}
\def\tL{\mathrm{L}}
\def\tp{\mathrm{p}}
\newcommand{\zbar}{\bar{z}}
\newcommand{\pd}[2]{\frac{\partial #1}{\partial #2}}
\newcommand{\F}[1]{\cF^{(#1)}}
\DeclareMathOperator{\Res}{Res}

\allowdisplaybreaks
\usepackage{rotating}
\newcommand{\be}{\begin{eqnarray}}
\newcommand{\beq}{\begin{eqnarray}}
\newcommand{\ee}{\end{eqnarray}}
\newcommand{\ov}{\overline}
\newcommand{\Ga}{\Gamma}

\newcommand{\ib}{\ov{i}}
\newcommand{\jb}{\ov{j}}

\newcommand{\ue}{\text{e}}
\newcommand{\h}{\frac{1}{2}}

\begin{document}

\setlength{\parindent}{0cm}
\setlength{\baselineskip}{1.5em}
\title{Topological Strings on Elliptic Fibrations}
\author{Murad Alim$^1$\footnote{\tt{alim@physics.harvard.edu}} and Emanuel Scheidegger$^2$\footnote{\tt{emanuel.scheidegger@math.uni-freiburg.de}}
\\
\small $^1$ Jefferson Physical Laboratory, Harvard University, \\ \small Cambridge, MA 02138, USA\\
\small $^2$Mathematisches Institut, Albert-Ludwigs-Universit\"at Freiburg,\\ \small Eckerstrasse 1, D-79104 Freiburg, Germany }

\date{}
\maketitle

\abstract{ We study topological string theory on elliptically fibered Calabi-Yau manifolds using mirror symmetry. We compute higher genus topological string amplitudes and express these in terms of polynomials of functions constructed from the special geometry of the deformation spaces. The polynomials are fixed by the holomorphic anomaly equations supplemented by the expected behavior at special loci in moduli space. We further expand the amplitudes in the base moduli of the elliptic fibration and find that the fiber moduli dependence is captured by a finer polynomial structure in terms of the modular forms of the modular group of the elliptic curve. We further find a recursive equation which captures this finer structure and which can be related to the anomaly equations for correlation functions.}

\clearpage


\tableofcontents
\section{Introduction}

Mirror symmetry and topological string theory are a rich source of insights in both mathematics and physics. The A- and B-model topological string theories probe K\"ahler and complex structure deformation families of two mirror Calabi-Yau (CY) threefolds $Z$ and $Z^*$ and are identified by mirror symmetry.  The B-model is more accessible to computations since its deformations are the complex structure deformations of $Z^*$ which are captured by the variation of Hodge structure. Mirror symmetry is established by providing the mirror maps which are a distinguished set of local coordinates in a given patch of the deformation space.  These provide the map to the A-model, since they are naturally associated with deformations of an underlying superconformal field theory and its chiral ring \cite{Lerche:1989uy}. 

At special loci in the moduli space, the A-model data provides enumerative information of the CY $Z$. This is contained in the Gromov-Witten invariants which can be resummed to give integer multiplicities of BPS states in a five-dimensional theory obtained form an M-theory compactification on $Z$ \cite{Gopakumar:1998ii,Gopakumar:1998jq}. Moreover, the special geometry governing the deformation spaces allows one to compute the prepotential $F_0(t)$ which governs the exact effective action of the four dimensional theories obtained from compactifying type IIA(IIB) string theory on $Z(Z^*)$ respectively.

The prepotential is the genus zero free energy of topological string theory, which is defined perturbatively in a coupling constant governing the higher genus expansion. The partition function of topological string theory indicating its dependence on local coordinates in the deformation space has the form:
\begin{equation}
\mathcal{Z}(t,\bar{t})=\exp \left(  \sum_{g} \lambda^{2g-2} \mathcal{F}^{(g)}(t,\bar{t}) \right) \, .
\end{equation} 
In refs.\cite{Bershadsky:1993ta,Bershadsky:1993cx}, Bershadsky, Cecotti, Ooguri and Vafa (BCOV) developed the theory and properties of the higher genus topological string free energies putting forward recursive equations, the holomorphic anomaly equations along with a method to solve these in terms of Feynman diagrams. For the full partition function these equations take the form of a heat equation \cite{Bershadsky:1993cx,Witten:1993ed} and can be interpreted \cite{Witten:1993ed} as describing the background independence of the partition function when the latter is interpreted as a wave function associated to the geometric quantization of $H^3(Z^*)$.

The higher genus free energies of the topological string can be furthermore interpreted as giving certain amplitudes of the physical string theory.\footnote{See ref.\cite{Antoniadis:2007ta} for a review.} The full topological string partition function conjecturally also encodes the information of $4d$ BPS states \cite{Ooguri:2004zv}. It is thus natural to expect the topological string free energies to be characterized by automorphic forms of the target space duality group. The modularity of the topological string amplitudes was used in \cite{Bershadsky:1993cx} to fix the solutions of the anomaly equation. The modularity of the amplitudes is most manifest whenever the modular group is $\tS\tL(2,\mZ)$ or a subgroup thereof. The higher genus generating functions of the Gromov-Witten invariants for the elliptic curve were expressed as polynomials \cite{Rudd:1994ta,Dijkgraaf:1995} where the polynomial generators were the elements of the ring of almost holomorphic modular forms $E_2,E_4$ and $E_6$ \cite{Kaneko:1995}. Polynomials of these generators also appear whenever $\tS\tL(2,\mZ)$ is a subgroup of the modular group, as for example in refs. \cite{Minahan:1998vr,Hosono:1999qc,Hosono:2002xj,Klemm:2004km}. The relation of topological strings and almost holomorphic modular forms was further explored in \cite{Aganagic:2006wq} (see also~\cite{Gunaydin:2006bz} and~\cite{Grimm:2007tm}).

Using the special geometry of the deformation space a polynomial structure of the higher genus amplitudes in a finite number of generators  was proven for the quintic and related one parameter deformation families \cite{Yamaguchi:2004bt} and generalized to arbitrary target CY manifolds \cite{Alim:2007qj}. The polynomial structure supplemented by appropriate boundary conditions enhances the computability of higher genus amplitudes. Moreover the polynomial generators are expected to bridge the gap towards constructing the appropriate modular forms for a given target space duality group which is reflected by the special geometry of the CY manifold.

In this work we use the polynomial construction to study higher genus amplitudes on elliptically fibered CY. The higher genus amplitudes are expressed in terms of a finite number of generators which are constructed from the special geometry of the moduli space of the CY.  Expanding the amplitudes of the elliptic fibration in terms of the base moduli allows us to further express the parts of the amplitudes depending on the fiber moduli in terms of the modular forms of $\tS\tL(2,\mZ)$. Together with this refinement of the polynomial structure we find a refined recursion which is the analog of an equation discovered in the context of BPS state counting of a non-critical string \cite{Minahan:1997ct,Minahan:1997if,Minahan:1998vr} and which was conjectured to hold for higher genus topological strings \cite{Hosono:1999qc,Hosono:2002xj}.

Writing the topological string amplitudes for the elliptic fibration which we consider in this work as an expansion:
$$F^{(g)}(t_E,t_B)= \sum_{n} f^{(g)}_n(t_E) q_B^n\, ,$$
where $t_E,t_B$ denote the special coordinates corresponding to the K\"ahler parameters of the fiber and base of the elliptic fibration respectively, $q_E=e^{2\pi i t_E},q_B=e^{2\pi i t_B}$, we find that $f^{(g)}_n$ can be written as 
$$f^{(g)}_n=P^{(g)}_n (E_2,E_4,E_6) \frac{q_E^{3n/2}}{\eta^{36n}}\, ,$$
where $P^{(g)}_n$ denotes a quasi-modular form constructed out of the Eisenstein series $E_2,E_4,E_6$ of weight $2g-2+18n$, we furthermore find the following recursion:

\begin{equation} \label{refinedrecursionIntro}
 \frac{\partial f^{(g)}_n}{\partial E_2}=-\frac{1}{24} \sum_{h=0}^g \sum_{s=1}^{n-1} s (n-s) f^{(h)}_s f^{(g-h)}_{n-s} +\frac{n(3-n)}{24} f^{(g-1)}_n \, .
\end{equation}

The outline of this work is as follows. In Section \ref{sec:mirrorsymmetry} we review some elements of mirror symmetry that allow us to set the stage for our discussion. We present and further develop techniques to identify the flat coordinates on the deformation spaces. In particular, we exhibit a systematic procedure to determine these coordinates at an arbitrary point in the boundary of the moduli space. 
We proceed in Section \ref{sec:recursion} with reviewing the holomorphic anomaly equations and how these can be used together with a polynomial construction to solve for higher genus topological string amplitudes. In Section \ref{sec:highergenus} we present the results of the application of the techniques and methods described earlier to an example of an elliptically fibered CY. The dependence on the moduli of the elliptic fiber can be further organized in terms of polynomials of $E_2,E_4$ and $E_6$ order by order in an expansion in the base moduli. We find a recursion~\eqref{refinedrecursionIntro} which captures this structure and relate it to the anomaly equation for the correlation functions of the full geometry. We show that such recursions hold for several examples of elliptic fibrations. We proceed with our conclusions in section \ref{conclusion}.

\section{Mirror symmetry}
\label{sec:mirrorsymmetry}
In this section we review some aspects of mirror symmetry which we will be using in the following.\footnote{See refs.\cite{Candelas:1990rm,Batyrev:1993dm,Hosono:1993qy} for foundational material as well as the review book \cite{Cox:1999ab} for general background on mirror symmetry. Some of the exposition in this section will follow refs.\cite{Alim:2009bx,Alim:2011rp}} To be able to fix the higher genus amplitudes we need a global understanding of mirror symmetry and how it matches expansion loci in the moduli spaces of the mirror manifolds $Z$ and $Z^*$.  We will also review and further develop some methods and techniques on the B-model side along refs.\cite{Lerche:1991wm,Ferrara:1991aj,Ceresole:1992su,Ceresole:1993qq,Hosono:1995bm,Hosono:1996jv,Noguchi:1996tu,Masuda:1998eh} to identify the special set of coordinates which allows an identification with the physical parameters and hence with the A-model side.


\subsection{Mirror geometries}
The mirror pair of CY 3-folds $(Z,Z^*)$ is given as hypersurfaces in toric ambient spaces $(W,W^*)$. The mirror symmetry construction of ref.\cite{Batyrev:1993dm} associates the pair $(Z,Z^*)$ to a pair of integral reflexive polyhedra $(\Delta,\Delta^*)$. 

\subsubsection*{\it The A-model geometry}
The polyhedron $\Delta^*$ is characterized by $k$ relevant integral points $\nu_i$ lying in a hyperplane of distance one from the origin in $\mZ^5$, $\nu_0$ will denote the origin following the conventions of refs. \cite{Batyrev:1993dm,Hosono:1993qy}.  The $k$ integral points $\nu_i(\Delta^*)$ of the polyhedron $\Delta^*$ correspond to homogeneous coordinates $u_i$ on the toric ambient space $W$ and satisfy $n=h^{1,1}(Z)$ linear relations:
\begin{equation}\label{toricrel}
\sum_{i=0}^{k-1} l_i^a \, \nu_i=0\, , \quad a=1,\dots,n\,.
\end{equation}
The integral entries of the vectors $l^a$ for fixed $a$ define the weights $l_i^a$ of the coordinates $x_i$ under the $\mC^*$ actions
$$ u_i \rightarrow (\lambda_a)^{l_i^a} u_i\,, \quad \lambda_a \in \mC^*\,.$$

The $l_i^a$ can also be understood as the $U(1)_a$ charges of the fields of the gauged linear sigma model (GLSM) construction associated with the toric variety \cite{Witten:1993yc}. The toric variety $W$ is defined as $W\simeq (\mC^{k}-\Xi)/(\mC^*)^n$, where $\Xi$ corresponds to an exceptional subset of degenerate orbits. To construct compact hypersurfaces, $W$ is taken to be the total space of the anti-canonical bundle over a compact toric variety. The compact manifold $Z \subset W$ is defined by introducing a superpotential $\mathcal{W}_Z=u_0 p(u_i)$ in the GLSM, where $x_0$ is the coordinate on the fiber and $p(u_i)$ a polynomial in the $u_{{i>0}}$ of degrees $-l_0^a$. At large K\"ahler volumes, the critical locus is at $u_0=p(u_i)=0$ \cite{Witten:1993yc}. 

An example of an elliptic fibration is the compact geometry given by a section of the anti-canonical bundle over the resolved weighted projective space $\mP(1,1,1,6,9)$. Mirror symmetry for this model has been studied in various places following refs.\cite{Hosono:1993qy,Candelas:1994hw}. The charge vectors for this geometry are given by:
\begin{equation}\label{chargevec}
\begin{array}{cccccccc}
&x_0&x_1&x_2&x_3&x_4&x_5&x_6\\
(l^1)=&(-6&3&2&1&0&0&0\, )\, ,\\
(l^2)=&(\hskip6pt 0&0&0&-3&1&1&1\, )\, .\\
\end{array}
\end{equation}

\subsubsection*{\it The B-model geometry}
The B-model geometry $Z^*\subset W^*$ is determined by the mirror symmetry construction of refs.\cite{Hori:2000kt,Batyrev:1993dm} as the vanishing locus of the equation
\begin{equation}
p(Z^*)=\sum_{i=0}^{k-1} a_i y_i =\sum_{\nu_i\in \Delta} a_i X^{\nu_i}\, ,
\end{equation}
where $a_i$ parameterize the complex structure of $Z^*$, $y_i$ are homogeneous coordinates \cite{Hori:2000kt} on $W^*$ and $X_m\, , m=1,\dots,4$ are inhomogeneous coordinates on an open torus $(\mC^*)^4 \subset W^*$  and $X^{\nu_i}:=\prod_m X_m^{\nu_{i,m}} $ \cite{Batyrev:1993wa}. The relations (\ref{toricrel}) impose the following relations on the homogeneous coordinates
\begin{equation}
\prod_{i=0}^{k-1} y_i^{l_i^a}=1\, ,\quad a=1,\dots,n=h^{2,1}(Z^*)=h^{1,1}(Z)\, .
\end{equation}
The important quantity in the B-model is the holomorphic $(3,0)$ form which is given by:
\begin{equation}\label{defomega0}
\Omega(a_i)= \textrm{Res}_{p=0} \frac{1}{p(Z^*)} \prod_{i=1}^4 \frac{dX_i}{X_i} \, .
\end{equation}
Its periods  $$\pi_{\alpha}(a_i)=\int_{\gamma^\alpha} \Omega(a_i)\, , \quad \alpha=0,\dots, 2h^{2,1}+1$$ are annihilated by an extended system of GKZ  \cite{Gelfand:1989} differential operators
\begin{eqnarray}
\mathcal{L}(l)= \prod_{l_i >0} \left( \frac{\partial}{\partial a_i}\right)^{l_i} -\prod_{l_i<0} \left( \frac{\partial}{\partial a_i}\right)^{-l_i}\, \\
\mathcal{Z}_k =\sum_{i=0}^{k-1} \nu_{i,j} \theta_i\, , \quad j=1,\dots,4\, . \quad \mathcal{Z}_0 = \sum_{i=0}^{k-1} \theta_i +1\,,\quad \theta_i=a_i \frac{\partial}{\partial a_i}\,,
\end{eqnarray}
where $l$ can be any positive integral linear combination of the charge vectors $l^a$. The equation $\mathcal{L}(l)\, \pi_0(a_i)=0$ follows from the definition (\ref{defomega0}). The equations $\mathcal{Z}_k\,\pi_\alpha(a_i)=0$ express the invariance of the period integral under the torus action and imply that the period integrals only depend on special combinations of the parameters $a_i$
\begin{equation}\label{lcs}
\pi_\alpha(a_i) \sim \pi_\alpha(z_a)\, ,\quad z_a=(-)^{l_0^a} \prod_i a_i^{l_i^a}\, ,
\end{equation}
the $z_a\,, a=1,\dots,n$ define local coordinates on the moduli space $\cM$ of complex structures of $Z^*$. 

In our example, there is an additional symmetry on $\cM$. Its origin is the fact that the polytope $\Delta^*$ has further integral points on facets~\cite{Hosono:1993qy,Candelas:1994hw}. They correspond to nonlinear coordinate transformations of the ambient toric variety $W$. These coordinate transformations can be compensated by transforming the parameters $a_i$. This yields the symmetry on $\cM$
\begin{equation}
  \label{eq:25}
  I\; : \; (z_1, z_2) \mapsto \left(\frac{1}{432}-z_1,-\frac{{z_1}^3z_2}{(\frac{1}{432}-z_1)^3}\right).
\end{equation}

The charge vectors defining the A-model geometry in Equ.(\ref{chargevec}) give the following Picard-Fuchs  (PF) operators annihilating $\tilde{\pi}_{\alpha}(z_i)=a_0\,\pi_{\alpha}(a_i)$:
\begin{eqnarray}\label{PF}
\mathcal{L}_1 &=& \theta_1 (\theta_1-3\theta_2) - 12 z_1 (6 \theta_1+1) (6 \theta_1+5)\, ,\\
 \mathcal{L}_2&=& \theta_2^3+z_2 \prod_{i=0}^2 (3\theta_2-\theta_1+i)\, ,\quad \theta_a:=z_a \frac{\partial}{\partial z_a} \, .
\end{eqnarray}
The discriminants of these operators are
\begin{equation}
  \label{eq:Discriminant}
  \begin{aligned}
    \Delta_1 &= (1- 432\,z_1)^3 - (432\,z_1)^3\,27\,z_2 ,\\
    \Delta_2 &= 1 + 27\, z_2 , 
  \end{aligned}
\end{equation}
Furthermore, we label the function 
\begin{equation}
  \label{eq:Delta_3}
  \Delta_3 = 1-432\,z_1.
\end{equation}
Note, that $I(\Delta_1) = (432\,z_1)^3\Delta_2$ and $I(\Delta_2) = \frac{\Delta_1}{{\Delta_3}^2}$, hence the vanishing loci of $\Delta_1$ and $\Delta_2$ are exchanged under the symmetry $I$.

A further important ingredient of mirror symmetry are the Yukawa
couplings which are identified with the genus zero correlators of
three chiral fields of the underlying topological field theory. In the B-model these are defined by:\footnote{We use $x^i\, ,i=1,\dots h^{2,1}$ to denote arbitrary coordinates on the moduli space of complex structures and denote a dependence on these collectively by $x$. We make the distinction to the coordinates defined in Equ.(\ref{lcs}) which will be identified with the coordinates centered around the large complex structure limiting point in the moduli space. }
\begin{equation}
C_{ijk}(x):= \int_{Z^*} \Omega \wedge \partial_i \partial_j \partial_k \Omega\, ,\quad \partial_i:=\frac{\partial}{\partial x^i}.
\end{equation}
For the example above these can be computed using the PF operators \cite{Hosono:1993qy}:
\begin{equation}
  \label{eq:Yukawa coupling 11169}
  \begin{aligned}
    C_{111}(z) &={\frac {9\,}{{z_{{1}}}^{3} \Delta_1 }} ,\\
    C_{112}(z) &= {\frac {3\, \Delta_3}{{z_{
{1}}}^{2} z_{{2}} \Delta_1}} ,\\
    C_{122}(z) &= {\frac { {\Delta_3} ^{2}}{z_{{1}}
{z_{{2}}}^{2} \Delta_1}} ,\\
   C_{222}(z) &= {\frac {9\, \left({\Delta_3}^3 + (432\,z_1)^3 \right)}{
       {z_{{2}}}^{2} \Delta_1 \Delta_2 }} .
  \end{aligned}
\end{equation}


\subsection{Variation of Hodge structure}
The Picard-Fuchs equations capture the variation of Hodge structure which describes the geometric realization on the B-model side of the deformation of the $\mathcal{N}=(2,2)$ superconformal field theory and its chiral ring \cite{Lerche:1991wm}, see also ref \cite{Ceresole:1993qq} for a review. Choosing one member of the deformation family of CY threefolds $Z^*$ characterized by a point in the moduli space $\cM$ of complex structures there is a unique holomorphic $(3,0)$ form $\Omega(x)$ depending on local coordinates in the deformation space. 

A variation of complex structure induces a change of the type of the reference $(3,0)$ form $\Omega(x)$. This change is captured by the variation of Hodge structure. $H^{3}(Z^*)$ is the fiber of a complex vector bundle over $\mathcal{M}$ equipped with a flat connection $\nabla$, the Gauss-Manin connection. The fibers of this vector bundle are constant up to monodromy of $\nabla$. The Hodge decomposition 
$$H^{3}=\bigoplus_{p=0}^3 H^{3-p,p}\, ,$$
varies over $\mathcal{M}$ as the type splitting depends on the complex structure. A way to capture this variation holomorphically is through the Hodge filtration $F^p$
\begin{equation}
H^3=F^0 \supset F^1\supset F^2\supset F^3\supset F^4=0\, , \quad F^p=\bigoplus_{q\ge p} H^{q,3-q} \subset H^3\, ,
\end{equation}
which define holomorphic subbundles $\mathcal{F}^p \to \cM$ whose fibers are $F^p$.  The Gauss-Manin connection on these subbundles has the property $\nabla \mathcal{F}^p \subset \mathcal{F}^{p-1} \otimes T^*\mathcal{M}$ known as Griffiths transversality.  This property allows us to identify derivatives of $\Omega(x) \in F^3$ with elements in the lower filtration spaces. The whole filtration can be spanned by taking multiderivatives of the holomorphic $(3,0)$ form. Fourth order derivatives can then again be expressed by the elements of the basis, which is reflected by the fact that periods of $\Omega(x)$ are annihilated the Picard-Fuchs system of differential equations of fourth order.
The dimensions of the spaces $(F^3,F^2/F^3,F^1/F^2,F^0/F^1)$ are $(1,h^{2,1},h^{2,1},1)$. Elements in these spaces can be obtained by taking derivatives of  $\Omega(x)$ w.r.t. the moduli.
For the example we are discussing a section of the filtration is given by the following vector $w(x)$ which has $2h^{2,1}+2=6$ components:
\begin{equation}
  \label{filtration}
w(x)= \left(\Omega(x),\, \quad \theta_1\Omega(x),\theta_2 \Omega(x),\quad \theta_1^2 \Omega(x),\theta_1\theta_2 \Omega(x),\quad \theta_1^2 \theta_2 \Omega(x)\, \right)^t\, .
\end{equation}
where $\theta_i = x^i \pd{}{x^i}$. Using $w(x)$ we can define the period matrix
\begin{equation}
\Pi(x)_{\beta}^{\phantom{\beta} \alpha}= \int_{\gamma^\alpha} w(x)\, ,\quad \gamma^{\alpha}\in H_3(Z^*)\, ,\quad\alpha,\beta=0,\dots,2 h^{2,1}+1\, ,
\end{equation}
the first row of which corresponds to the periods of $\Omega(x)$. The periods are annihilated by the PF operators. We can identify solutions of the PF operators 
with the periods of $\Omega(x)$. In our example, near the point of maximal unipotent monodromy $z=(z_1,z_2)$, the solutions are given in the Appendix~\ref{sec:gaussmanin}.

\subsubsection*{{\it Polarization}}
The variation of Hodge structure of a family of Calabi--Yau threefolds in addition comes with a polarization, i.e. a nondegenerate integral bilinear form $Q$ which is antisymmetric. This form is defined by $Q(\varphi,\psi) = \int_{Z^*} \varphi \wedge \psi$ for $\varphi, \psi \in H^3$. The polarization satisfies \[ Q(F^p , F^{4-p} ) = 0, \qquad Q(C \varphi, \bar\varphi ) > 0 \textrm{ for } \varphi \ne 0,\]
where $C$ acts by multiplication of $i^{p-q}$ on $H^{p,q}$. Hence, $Q$ is a symplectic form.

Since the space of periods can be identified with the space of
solutions to the Picard--Fuchs equations, the symplectic form on $H^3(Z^*)$ should be expressible in terms of a bilinear operator acting on the space of solutions. This approach has been developed in ref.\cite{Masuda:1998eh}. We will review and employ these techniques in the following. 

We want to express the symplectic form $Q$ in terms of the
basis~(\ref{filtration}). For this purpose, we define an antisymmetric
linear bidifferential operator on the space of solutions of the Picard--Fuchs equation as
  \begin{equation}
    \label{eq:77}
    D_1 \wedge D_2(f_1, f_2) = \frac{1}{2}\left( D_1f_1D_2f_2 - D_2f_1D_1f_2\right)\,,
  \end{equation}
where $D_1$ and $D_2$ are arbitrary differential operators with respect to $x$. Then we can write $Q$ as an antisymmetric bidifferential operator 
\begin{equation}
  \label{eq:AnsatzforQ}
  Q(x) = \sum_{k,l} Q_{k,l}(x) D_k(\theta) \wedge D_l(\theta)\,,
\end{equation}
where $D_k, D_l$ run over the basis~(\ref{filtration}) of multi-derivatives used to define the vector $w(x)$ spanning the Hodge filtration. The condition that
$Q(x)$ is constant over the moduli space, i.e. 
\begin{equation}
  \label{eq:ConstancyofQ}
  \theta_i Q(x) = 0, \qquad i=1,\dots, h^{2,1},
\end{equation}
imposes constraints on the coefficients $Q_{k,l}(x)$. These lead to a
system of algebraic and differential equations for the
$Q_{k,l}(x)$. At this point we need to express the higher order
differential operators in terms of the basis~(\ref{filtration}) using the
relations such as~(\ref{eq:Degree2relation}) and~(\ref{eq:Degree3relation}). 
Then this system can be solved up to an overall constant.

In our example, near the point of maximal unipotent monodromy $z=(z_1,z_2)$, we find
\begin{equation}
  \label{eq:Q}
  \begin{aligned}
    Q(z) =&\frac{1}{3}\,\Delta_2\Delta_3\left( \theta_{{1}} \wedge {\theta_{{2}}}^{2} + 
\theta_{{2}} \wedge \theta_{{1}}\theta_{{2}}\right) - \Delta_2\, \theta_{{2}} \wedge
{\theta_{{2}}}^{2}-{\frac {a_9 }{3\, \Delta_3}}\, \theta_{{1}} \wedge \theta_{{1}}\theta_{{2}}\\
    &-{\frac {\Delta_1 }{ 3\, {\Delta_3} ^{2}}}\,1\wedge \theta_{{1}} {\theta_{{2}}}^{2} +{\frac {a_{10}}{{\Delta_3}  ^{2}}}\,1 \wedge\theta_{{1}}\theta_{{2}}+{\frac { a_4 }{ 3\, {\Delta_3} ^{2}}}\,1 \wedge \theta_{{1}} + {\frac { 20\, z_{{1}}a_9}{ {\Delta_3} ^{2}}}\,1 \wedge\theta_{{2}}.
  \end{aligned}
\end{equation}
where $a_4$, $a_9$ and $a_{10}$ are given in~(\ref{eq:as}). In the basis of periods~(\ref{eq:22}) we then obtain
\begin{equation}
  \label{eq:Qmatrix}
  \left( \begin {array}{cccccc} 0&0&0&0&0&1/2\\ \noalign{\medskip}0&0&0
&0&-1/2&0\\ \noalign{\medskip}0&0&0&-1/2&0&0\\ \noalign{\medskip}0&0&1/2
&0&0&0\\ \noalign{\medskip}0&1/2&0&0&0&0\\ \noalign{\medskip}-1/2&0&0
&0&0&0\end {array} \right)
\end{equation}

Moreover, the invariant definition of the B--model prepotential is given in
terms of the natural symplectic form $Q$ on $H^3(Z^*,\mZ)$. Let
$\varpi_i(x)$ be a basis for the periods, then 
\begin{equation}
  \label{eq:Prepotential}
  \cF^{(0)}(x) = \frac{1}{2} \sum_{i>j} Q(\varpi_i(x), \varpi_j(x))\,.
\end{equation}


\subsection{The Gauss-Manin connection and flat coordinates}
\label{sec:periods-mirror-map}
\subsubsection*{\it The Gauss-Manin connection}
The Picard-Fuchs operators (\ref{PF}) are equivalent to a first order equation for the period matrix. Using linear combinations of the operators and derivatives thereof, the system can be cast in the form
\begin{equation}
  \label{eq:Gauss-Manin}
  \left(\theta_i- A_i(x) \right) \Pi(x)_{\beta}^{\phantom{\beta} \alpha} =0\, ,\quad i=1,\dots,h^{2,1}\, ,
\end{equation}
which defines the Gauss-Manin connection $\nabla$. For our example, the matrices $A_i(x)$ near the point of maximal unipotent monodromy are given in the appendix.

There are limiting points in the moduli space of complex structure $\cM$ at which the Hodge structure degenerates~\cite{Deligne:1970ab,Cox:1999ab}. These points are of particular interest in the expansion of the topological string amplitudes. In order to describe these limiting points, we assume that there exists a smooth compactification $\overline \cM$ of $\cM$ such the boundary consists of a finite set $I$ of normal crossing divisors $\overline\cM \setminus \cM = \bigcup_{i \in I} D_i$. Along these divisors, the Gauss--Manin connection can acquire regular singularities. This means that, at a point $p \in \bigcap_{i=1}^{h^{2,1}} D_i$, the connection matrix has at worst a simple pole along $D_i$. Note that since we defined $A_i$ in~(\ref{eq:Gauss-Manin}) with $\theta_i$ instead of $\partial_i$ this means that matrix $A_i(z)$ is holomorphic along $D_i$. 

At a regular singularity described by a divisor $D_i = \{y_i=0\}$\footnote{We will denote local coordinates near an intersection point of boundary divisors by $y$, still reserving $z$ for the point of maximal unipotent monodromy.} we therefore define:
\begin{equation}
\label{eq:83}
  \Res_{D_i}(\nabla) = A_i(y)|_{y_i=0} .
\end{equation}
This residue matrix gives the following useful information. The eigenvalues of the monodromy $T$ are $\exp(2\pi i \lambda)$ as $\lambda$ ranges over the eigenvalues of $\Res(\nabla)$. Furthermore, $T$ is unipotent if and only if $\Res(\nabla)$ has integer eigenvalues. Finally, if no two distinct eigenvalues of $\Res(\nabla)$ differ by an integer, then $T$ is conjugate to $S = \exp(-2\pi i \Res(\nabla))$. These properties allow us to extract the relevant information about the monodromy of $\nabla$ around these boundary divisors. We will see later that this allows us to easily obtain the solutions to the Picard--Fuchs equations at the various boundary points.

The monodromies $T_i$ for all the divisors $D_i$ in the boundary form a group, the monodromy group  $\Gamma$ of the Gauss--Manin connection. This group is a subgroup of $\textrm{Aut}(H^3(Z^*,\mZ))$ preserving the symplectic form $Q$. Hence, $\Gamma$ is a subgroup of $\tS\tp(2h^{2,1} + 2,\mZ)$. The topological string amplitudes $\F{g}$ are expected to be automorphic with respect to this group. 

The point $p$ in the boundary which has been studied usually so far, is the point of maximal unipotent monodromy, also known as the large complex structure limit.
From the connection matrices $A_i(x)$ of our example we can immediately get information on the monodromy matrices around the divisors $D_{(1,0)} = \{z_1 = 0\}$ and $D_{(0,1)} = \{z_2 = 0\}$. (For the notation on the divisors see Section~\ref{sec:moduli-space-its}.) We simply consider the matrices $\Res_{\{z_i=0\}} = A_i(z)|_{z_i=0}$ and bring them into Jordan normal form. This yields
\begin{equation}
  \label{eq:Residue}
  \begin{aligned}
    \Res_{D_{(1,0)}}(\nabla) &\sim
\left( \begin {array}{cccccc} 0&1&0&0&0&0\\ \noalign{\medskip}0&0&1&0
&0&0\\ \noalign{\medskip}0&0&0&1&0&0\\ \noalign{\medskip}0&0&0&0&0&0
\\ \noalign{\medskip}0&0&0&0&0&0\\ \noalign{\medskip}0&0&0&0&0&0
\end {array} \right), &
    \Res_{D_{(0,1)}}(\nabla) &\sim
\left( \begin {array}{cccccc} 0&1&0&0&0&0\\ \noalign{\medskip}0&0&1&0
&0&0\\ \noalign{\medskip}0&0&0&0&0&0\\ \noalign{\medskip}0&0&0&0&1&0
\\ \noalign{\medskip}0&0&0&0&0&1\\ \noalign{\medskip}0&0&0&0&0&0
\end {array} \right)     
  \end{aligned}
\end{equation}
From this we read off that the corresponding monodromy matrices $T_{D_{(1,0)}}$ and $T_{D_{(0,1)}}$ satisfy
\begin{equation}
  \label{eq:Unipotency}
  \begin{aligned}
    \left(T_{D_{(1,0)}} - 1\right)^4 &=0, & \left(T_{D_{(0,1)}} - 1\right)^3 &=0. 
  \end{aligned}
\end{equation}
It can be checked that these monodromy matrices satisfy the conditions for a point of maximal unipotent monodromy~\cite{Candelas:1994hw,Cox:1999ab}.

\subsubsection*{\it Flat coordinates}
\label{sec:flat-coordinates}
We proceed by discussing a special set of coordinates on the moduli space of complex structure which permit an identification with the physical deformations of the underlying theory. These coordinates are defined within special geometry which was developed studying moduli spaces of $\mathcal{N}=2$ theories, we follow refs. \cite{Lerche:1989uy,Candelas:1990pi,Candelas:1990rm,Strominger:1990pd,Lerche:1991wm,Ferrara:1991aj,Ceresole:1992su,Ceresole:1993qq}. 
Choosing a symplectic basis of 3-cycles $A^I,B_J \in H_3(Z^*)$ and a dual basis $\alpha_I,\beta^J$ of $H^3(Z^*)$ such that 
\begin{eqnarray}
&&A^I\cap B_J=\delta^I_J=-B_J\cap A^I\,,\quad  A^I\cap A^J=B_I\cap B_J=0\, \, ,\nonumber\\
&& \int_{A^I} \alpha_J =\delta^I_J\, , \quad\int_{B_J} \beta^I =\delta^I_J\, ,\quad  I,J=0,\dots h^{2,1}(Z^*)\, ,
\end{eqnarray}
the $(3,0)$ form $\Omega(x)$ can be expanded in the basis $\alpha_I,\beta^J$:
\begin{equation}
\Omega(x)= X^I(x) \alpha_I - \mathcal{F}_J(x) \beta^J\, .
\end{equation}
The periods $X^I(x)$ can be identified with projective coordinates on
$\mathcal{M}$ and $\mathcal{F}_J$ with derivatives of a function
$\mathcal{F}(X^I)$, $\mathcal{F}_J=\frac{\partial
  \mathcal{F}(X^I)}{\partial X^J}$. In a patch where $X^0(x)\ne 0$ a
set of special coordinates can be defined $$t^a=\frac{X^a}{X^0}\,
,\quad a=1,\dots ,h^{2,1}(Z^*).$$ The normalized holomorphic $(3,0)$
form $v_0= (X^0)^{-1} \Omega(t)$ has the expansion:
\begin{equation}
 v_0= \alpha_0 + t^a \alpha_a -\beta^b F_b(t) - (2F_0(t)-t^c F_c(t)) \beta^0\,,
\end{equation}
where $$F_0(t)= (X^0)^{-2} \mathcal{F} \quad \textrm{and} \quad F_a(t):=\partial_a F_0(t)=\frac{\partial F_0(t)}{\partial t^a}.$$
$F_0(t)$ is the prepotential. We define further 
\begin{eqnarray}
v_a &=& \alpha_a-\beta^b F_{ab}(t)-(F_a(t)-t^bF_{ab}(t))\beta^0\, ,\\
v_D^a &=& -\beta^a -t^a \beta^0\, ,\\
v^0 &=& \beta^0\, .
\end{eqnarray}
The Yukawa coupling in special coordinates is given by
\begin{equation}
C_{abc}:= \partial_a \partial_b \partial_c F_0(t)=\int_{Z^*} v_0 \wedge \partial_a\partial_b\partial_c v_0\, .
\end{equation}
We further define the vector with $2h^{2,1}+2$ components:
\begin{equation}
v=(v_0\,, \quad v_a\,,\quad v_D^a\,,\quad v^0)^t\,,
\end{equation}

We have then by construction:
\begin{equation}
\partial_a \left( \begin{array}{c} v_0\\v_b\\v_D^b\\v^0 \end{array}\right)=\underbrace{\left( \begin{array}{cccc}
0&\delta^c_a&0&0\\
0&0&C_{abc}&0\\
0&0&0&\delta^b_a\\
0&0&0&0
 \end{array}
\right)}_{:=C_a} \, \left( \begin{array}{c} v_0\\v_c\\v_D^c\\v^0 \end{array}\right)\,,
\end{equation}
which defines the $(2h^{2,1}+2) \times (2h^{2,1}+2)$ matrices $C_a$, in terms of which we can write the equation in the form:
\begin{equation}\label{GaussManin}
\left(\partial_a-C_a\right) \, v=0\, .
\end{equation}

The entries of $v$ correspond to elements in the different filtration spaces discussed earlier. As in~\eqref{eq:Gauss-Manin}, Equ.(\ref{GaussManin}) defines the Gauss-Manin connection, now in special coordinates. The upper triangular structure of the connection matrix reflects the effect of the charge increment of the elements in the chiral ring upon insertion of a marginal operator of unit charge. Since the underlying superconformal field theory is isomorphic for the A- and the B-models, this set of coordinates describing the variation of Hodge structure is the good one for describing mirror symmetry and provide thus the mirror maps. The following discussion builds on refs.\cite{Hosono:1995bm,Hosono:1996jv,Noguchi:1996tu}.

In order to find the mirror maps starting from a set of arbitrary local coordinates on the moduli space of complex structure we study the relation between the vectors $w$ of Equ.(\ref{filtration}) and $v$ spanning the Hodge filtration, these are related by the following change of basis:
\begin{equation}
  \label{eq:Change_basis}
  w(x(t)) = M(x(t)) v(t)\,.
\end{equation}
By the fact that this change of basis is filtration-preserving, the matrix
$M(x)$ must be lower block-triangular. For concreteness we expose the discussion in the following for $h^{2,1}(Z^*)=2$:
\begin{equation}
  \label{eq:M}
  M(x) = 
    \begin{pmatrix}
      m_{11} & 0 & 0 & 0 & 0 & 0\\
      m_{21} & m_{22} & m_{23} & 0 & 0 & 0\\
      m_{31} & m_{32} & m_{33} & 0 & 0 & 0\\
      m_{41} & m_{42} & m_{43} & m_{44} & m_{45} & 0\\
      m_{51} & m_{52} & m_{53} & m_{54} & m_{55} & 0\\
      m_{61} & m_{62} & m_{63} & m_{64} & m_{65} & m_{66}
   \end{pmatrix}
\end{equation}
Imposing that the change of connection matrices yields the desired result requires the vanishing of the following matrix:
\begin{equation}
  N_a(t) = C_a(t) - \sum_i\, J_{ia} \, M(x)^{-1}\left(A_i(x)M(x) -
      \theta_i M(x) \right)\,.\\
  \label{eq:N_i}
\end{equation}
Here $J= (J_{ia})$ is the Jacobian for the logarithmic derivative
\begin{equation}
  \label{eq:24}
  J_{ia} = \frac{1}{x^i}\pd{x^i}{t^a} .
\end{equation}
The matrices $N_a$ have the general block form
\begin{equation}
  \label{eq:60a}
  N_a(x) = 
    \begin{pmatrix}
      n_{a,11} & n_{a,12} & n_{a,13} & 0 & 0 & 0\\
      n_{a,21} & n_{a,22} & n_{a,23} & n_{a,24} & n_{a,25} & 0\\
      n_{a,31} & n_{a,32} & n_{a,33} & n_{a,34} & n_{a,35} & 0\\
      n_{a,41} & n_{a,42} & n_{a,43} & n_{a,44} & n_{a,45} & n_{a,46}\\
      n_{a,51} & n_{a,52} & n_{a,53} & n_{a,54} & n_{a,55} & n_{a,56}\\
      n_{a,61} & n_{a,62} & n_{a,63} & n_{a,64} & n_{a,65} & n_{a,66}
   \end{pmatrix}
\end{equation}

We set $m_{11}(x)=X^0(x)$ since it will turn out that this quantity should be identified with one of the periods. The vanishing of the first column of the $N_a$ allows us to express the $m_{k1}$ in terms of $X^0(x)$ and its derivatives. Moreover, it follows that $m_{11}$ is a solution to the Picard--Fuchs equations
\begin{equation}
  \label{eq:1}
  \cL_r\,  X^0(x) = 0\,, 
\end{equation}
Similarly, the vanishing of the second and third column of the $N_a$ expresses the $m_{k2}$ and $m_{k3}$ in terms of $m_{12}$ and $m_{13}$ and their derivatives, respectively. In addition, they satisfy differential equations of the form
\begin{equation}
  \label{eq:2}
  \cD_r (t_a\, X^0) = \cL_r ( t_a\, X^0) - t_a \cL_r\, X^0 = 0.
\end{equation}
Together with~(\ref{eq:1}) we conclude that the products $t_1 X^0$ and $t_2 X^0$ must be solutions to the Picard--Fuchs equations as well. In other words, the flat coordinates must be ratios of two periods. The differential equations~(\ref{eq:2}) form a system of nonlinear partial differential equation which determine the flat coordinates in terms of $x$. In general, they are hard to solve, but one can transform this system into a system of linear partial differential equations of higher order along the lines of~\cite{Klemm:1994wn}.

Next, we consider the blocks $\left(\begin{smallmatrix} n_{a,24} & n_{a,25}\\n_{a,34} & n_{a,35} \end{smallmatrix}\right)=0$. They can be solved for the functions $C_{abc}(t)$. This yields expressions in terms of $t_a$, their derivatives, and the functions $m_{22}$, $m_{23}$, $m_{32}$, $m_{33}$, $m_{44}$, $m_{45}$, $m_{54}$, $m_{55}$. Taking into account the previous results, we need to express the latter four functions in terms of $X^0$. 

The two conditions $n_{a,46}=0$ can be used to express $m_{44}$ and $m_{45}$ in terms of $t_a$, their derivatives, and $m_{66}$. Similarly, $n_{a,56}=0$ yield similar expression for $m_{54}$ and $m_{55}$. If we apply this to our example and again choose the point of maximal unipotent monodromy with local coordinates $z$, then we obtain the following relations
\begin{equation}
  \label{eq:5}
  \begin{aligned}
    m_{44}(z) &= \frac{ 3\,\theta_2 t_2 - \Delta_3 \theta_1 t_2}{ \Delta_3 \det J} m_{66}(z), \\ 
    m_{45}(z) &= -\frac{ 3\,\theta_2 t_1 - \Delta_3 \theta_1 t_1}{ \Delta_3 \det J} m_{66}(z),\\ 
    m_{54}(z) &= -\frac{\left(9-11664\,z_{{1}}+5038848\,{z_{{1}}}^{2}\right) \theta_1t_2 - {\Delta_3}^2\Delta_2 \theta_2t_2}{{\Delta_3}^2 \Delta_2 \det J} m_{66}(z),\\  
   m_{55}(z) &= \frac{\left(9-11664\,z_{{1}}+5038848\,{z_{{1}}}^{2}\right) \theta_1t_1 - {\Delta_3}^2\Delta_2 \theta_2t_1}{{\Delta_3}^2 \Delta_2 \det J} m_{66}(z).
  \end{aligned}
\end{equation}
The vanishing of $n_{1,44}$ and $n_{1,45}$ allows to express $m_{64}$ and $m_{65}$ in terms of $m_{42},\dots,m_{45}$, $t_a$, their derivatives and the $C_{abc}$. Upon using the previous results, they can be expressed in terms of $X^0$, $t_i$, their derivatives, and $m_{66}(z)$. 

To determine the latter, we use the vanishing of the $n_{a,66}$.
\begin{equation}
  \label{eq:6}
  {\frac {432\, z_{{1}} \left( \Delta_1+30233088\,{z_{{1}}}^{2}z_{{
2}} \right)  }{ \Delta_1 \Delta_3 }} m_{{6,6}}(z)  -\theta_1 
 m_{{6,6}}(z)  -(\theta_1t_1)m_{64}(z) -(\theta_1t_2)m_{65}(z) = 0.
\end{equation}
Substituting all the previous results leads to the following differential equation
\begin{equation}
  \label{eq:7}
  \Delta_1\Delta_3\left(m_{66}(z)\theta_1X^0(z)+X^0(z)\theta_1m_{66}(z)\right) - 432\,z_1 \left( \Delta_1+30233088\,{z_{{1}}}^{2}z_{{
2}} \right) m_{66}(z)X^0(z) = 0.
\end{equation}
All the dependence on the $t_i$ has cancelled. We observe that the prefactor of $m_{66}(z)X^0(z)$ can be written as
\begin{equation}
  \label{eq:8}
  \frac{{\Delta_1}^2}{\Delta_3}\theta_1 \left( \frac{{\Delta_3}^2}{\Delta_1} \right) = 432\,z_1 \left( \Delta_1+30233088\,{z_{{1}}}^{2}z_{{
2}} \right).
\end{equation}
Hence, the differential equation simplifies to
\begin{equation}
  \label{eq:9}
  \theta_1 \left(\frac{{\Delta_3}^2}{\Delta_1 m_{66}(z) X^0(z)}\right) = 0.
\end{equation}
Its solution is
\begin{equation}
  \label{eq:10}
  m_{66}(z) = f(z_2) \frac{{\Delta_3}^2}{\Delta_1 X^0(z)},
\end{equation}
where $f(z_2)$ is an undetermined function that only depends on $z_2$. To fix this function we look at the vanishing of the $n_{2,66}$. After all substitutions this yields the differential equation
\begin{equation}
  \label{eq:11}
  \theta_2( \Delta_1 m_{66}(z) X^0(z) ) = \theta_2\left(f(z_2) {\Delta_3}^2\right) = 0.
\end{equation}
Since $\Delta_3$ does not depend on $z_2$, we conclude that $f(z_2)$ must be a constant, which we set to 1. 

We can now recursively express all the functions $m_{ij}$ through the function $X^0(z)$ which must be a solution of the Picard--Fuchs equations. In particular, this yields the well known expression for the Yukawa couplings in flat coordinates
\begin{equation}
  \label{eq:4}
  C_{abc}(t) = \sum_{i,j,k} \frac{1}{\left(X^0(z(t))\right)^2} \pd{z_i}{t^a}\pd{z_j}{t^b}\pd{z_k}{t^c} C_{ijk}(z(t))\,.
\end{equation}

There are still a few conditions remaining, namely $n_{a,64}=0$ and $n_{a,65}=0$. These turn out to be very difficult to analyze. One can check that these conditions are implied by
\begin{align}
  \label{eq:16}
  Q( X^0, t_1 X^0) &= 0, & Q( X^0, t_2 X^0) &= 0, & Q( t_1 X^0, t_2 X^0) &= 0.
\end{align}
where $Q$ was determined in~(\ref{eq:Q}). In particular, not every ratio of solutions to the Picard--Fuchs equations yields a flat coordinate. In general, we expect a weaker condition involving the left-hand sides of~(\ref{eq:16}) to be equivalent to the vanishing of $N_a$.

\subsubsection*{{\it Solutions of the Picard--Fuchs equations}}
\label{sec:solut-picard-fuchs}

As we have just seen, in order to determine the flat coordinates we need solutions of the Picard--Fuchs equation which satisfy~(\ref{eq:16}). It is well-known how to solve these equations at the point of maximal unipotent monodromy by observing that they form extended GKZ hypergeometric systems, see e.g.~\cite{Hosono:1993qy,Hosono:1994ax}. However, we will need the flat coordinates at other special loci in the moduli space. For this purpose we need a systematic procedure to solve the system of Picard--Fuchs equations at an arbitrary point in the boundary $\overline\cM \setminus \cM$ of the moduli space where it is in general no longer of extended GKZ hypergeometric type. 

However, if the moduli space $\cM$ is one-dimensional we have the following well-known result, see e.g.~\cite{Coddington:1955ab,Borel:1987ab}. Let \[R = \Res_{y=0} \nabla = A(y)|_{y=0}\] be the residue matrix of the connection $\nabla$ at a regular singular point given by $y=0$. $R$ is a  constant matrix. Then there exists a fundamental system of solutions to~(\ref{eq:Gauss-Manin}) of the form \[u(y) = y^R S(y)\] with $S(y)$ a single-valued and holomorphic matrix. Since any two fundamental systems are related by an invertible constant matrix, this form is independent of the choice of basis, and we can take for $R$ its Jordan normal form. This simplifies the computations enormously. 

In the present case where the moduli space $\cM$ is higher-dimensional we can prove the following result: Let $p = \bigcap_{i=1}^{n} D_i$ be a point at the intersection of $h^{2,1}$ boundary divisors, where each of the divisors $D_i$ is given by an equation $y_i=0$. Let \[R_i = \Res_{D_i}\nabla = A_i(y)|_{y_i=0}, \quad \forall i.\] 
The matrices $R$ are in general not constant anymore. Then a fundamental system of solutions takes the form \[ u(y) = \prod_{i=1}^{n} {y_i}^{R_i} S(y) . \] This follows by induction from the result in dimension 1 together with the fact that $[R_i, R_j] = 0$, a consequence of the flatness of $\nabla$. Moreover, if $J_i$ is the Jordan normal form of $R_i$, then there exist constant matrices $P_i$ such that \[ u(y) = \prod_{i=1}^n P_i\, {y_i}^{J_i} \, S(y) . \] This form considerably simplifies the explicit computation. In practice, the $P_i$ are often permutation matrices.

\subsubsection*{{\it Elliptic fibrations}}

Here, we discuss a few aspects of elliptic fibrations. Let $Z$ be an elliptically fibered Calabi--Yau threefold $\pi: Z \to B$ where the fiber $\pi^{-1}(p) \cong E$ is a smooth elliptic curve, $p \in B \setminus \Delta$, where the discriminant $\Delta$ is a divisor in $B$.  We consider the variation of Hodge structure for the family of mirror
Calabi--Yau threefolds $f:\cZ^* \to \cM$ where $\cM$ is the complex
structure moduli space. We recall that the Gauss--Manin connection for this family
has monodromy group $\Gamma \in \textrm{Aut}(H^3(Z^*,\mZ))$. Since $Z$ is an elliptic fibration, there is a distinguished subgroup of $\Gamma$ isomorphic to a subgroup $\Gamma_{\text{ell}} \subset \tS\tL_2(\mZ)$ and the variation of Hodge structure contains a variation of sub--Hodge structures coming from the elliptic fiber. 

In our example the monodromy group $\Gamma$ is generated by two matrices
$A$ and $T$~\cite{Candelas:1994hw}. Consider the element
$T_\infty = (TA)^{-1} \in \Gamma$. Then $A^3$ and ${T_\infty}^3$
generate an $\tS\tL_2(\mZ)$ subgroup as follows:
\begin{equation}
  \label{eq:SL2Z}
  \begin{aligned}
    A^3 t_1 &= -\frac{1}{t_1+1}, & {T_\infty}^3 t_1 &= t_1 + 1
  \end{aligned}
\end{equation}
Hence, we expect $t_1$ to be a modular parameter of an elliptic
curve. In fact, in the limit $z_2 \to 0$ and the Picard--Fuchs system reduces to the Picard--Fuchs equation of the elliptic curve mirror to the elliptic fiber.

  
\section{Higher genus recursion}
\label{sec:recursion}
In this section we review the ingredients of the polynomial
construction \cite{Yamaguchi:2004bt,Alim:2007qj}, following
\cite{Alim:2007qj} as well as the boundary conditions needed to
supplement the construction to fix remaining ambiguities. To implement
the boundary conditions it is necessary to be able to provide the good
physical coordinates in every patch in moduli space. This can be done by exploiting the flat structure of the variation of Hodge structure on the B-model side.

\subsection{Special geometry and the holomorphic anomaly}
The deformation space $\mathcal M$ of topological string theory, parameterized by coordinates
$x^i$, $i=1,...,\textrm{dim}(\mathcal{M})$, carries the structure of a special K\"ahler manifold.\footnote{See ref.\cite{Bershadsky:1993cx} for background material.} The ingredients of this structure are the Hodge line bundle $\mathcal{L}$ over $\mathcal M$ and the cubic couplings which are a holomorphic section of  $\mathcal{L}^{2} \otimes \textrm{Sym}^3 T^* \mathcal{M}$. The metric on $\mathcal{L}$ is denoted by $e^{-K}$ with respect to some local trivialization and provides a K\"ahler potential for the special K\"ahler metric on $\mathcal{M}$, $G_{i \jb}=\partial_i \bar{\partial}_{\jb}K$. Special geometry further gives the following expression for the curvature  of $\mathcal{M}$ 
\begin{equation}
R_{i\ib\phantom{l}j}^{\phantom{i\ib}l}=[\bar{\partial}_{\ib},D_i]^l_{\phantom{l}j}=\bar{\partial}_{\bar{i}} \Gamma^l_{ij}= \delta_i^l
G_{j\bar{i}} + \delta_j^l G_{i\bar{i}} - C_{ijk} \bar C^{kl}_{\bar{i}}.
\label{curvature}
\end{equation}
The topological string amplitude or partition function $\F{g}$ at genus $g$ is a section of the line bundle
$\mathcal{L}^{2-2g}$ over $\mathcal M$. The correlation function at genus $g$ with $n$
insertions $\mathcal{F}^{(g)}_{i_1\cdots i_n}$ is only non-vanishing for
$(2g-2+n)>0$. They are related by taking covariant derivatives as this
represents insertions of chiral operators in the bulk, e.g.~$D_i \mathcal{F}^{(g)}_{i_1\cdots i_n}=\mathcal{F}^{(g)}_{ii_1\cdots i_n}$.

$D_i$ denotes the covariant derivative on the bundle $\cL^m\otimes 
\textrm{Sym}^n T^*\mathcal{M}$ where $m$ and $n$ follow from the context. \footnote{The notation $D_i$ is also being used for the boundary divisors $D_i \in \overline\cM \setminus \cM$. It is clear from the context which meaning applies.} $T^*\mathcal{M}$ is
the cotangent bundle of $\cM$ with the standard connection coefficients
$\Ga^i_{jk}=G^{i\ib}\partial_jG_{k\ib}$. The connection on the bundle $\cL$ is
given by the first derivatives of the K\"ahler potential $K_i=\partial_iK$.

In \cite{Bershadsky:1993cx} it is shown that the genus $g$ amplitudes
are recursively related to lower genus amplitudes by the holomorphic anomaly equations:
\begin{eqnarray}
\bar{\partial}_{\bar{i}} \F{g}_{i_1\dots i_n} &=& \frac{1}{2} \bar{C}_{\bar{i}}^{jk} \left(
\sum_{r=0}^{g}\sum_{s=0}^{n} \frac{1}{s!(n-s)!} \sum_{\sigma \in S_n}
D_j\mathcal{F}^{(r)}_{i_{\sigma(1)} \dots i_\sigma(s)} D_k\mathcal{F}^{(g-r)}_{i_{\sigma(s+1)}\dots i_{\sigma(n)}} +
D_jD_k\mathcal{F}^{(g-1)}_{i_1\dots i_n} \right),\nonumber\\
&&-(2g-2+n-1)\sum_{s=1}^n G_{\bar{i}i_s} \mathcal{F}^{g}_{i_1\dots i_{s-1}i_s\dots i_n}\, ,
\end{eqnarray}
where
\begin{equation}
\bar{C}_{\bar{k}}^{ij}= \bar{C}_{\bar{i} \bar{j}\bar{k}} G^{i
\bar{i}}G^{j \bar{j}}\, \ue^{2K}, \qquad \bar{C}_{\bar{i}\bar{j}\bar{k}}=
\overline{C_{ijk}}\,. \label{Cbar}
\end{equation}
and where the sum $\sigma \in S_n$ is over permutations of the insertions and the formula is valid for $(g=0,n\ge4)$, $(g=1,n\ge2)$ and all higher genera and number of insertions. For $n=0$ it reduces to the holomorphic anomaly for the free energies $\mathcal{F}^{g}$:
\begin{equation}
\bar{\partial}_{\bar{i}} \F{g} = \h \bar{C}_{\bar{i}}^{jk} \left(
\sum_{r=1}^{g-1}
D_j\mathcal{F}^{(r)} D_k\mathcal{F}^{(g-r)} +
D_jD_k\mathcal{F}^{(g-1)} \right) \label{anom1}.
\end{equation}
These equations, supplemented by \cite{Bershadsky:1993ta}
\begin{equation}
\bar{\partial}_{\bar{i}} \mathcal{F}^{(1)}_j = \frac{1}{2} C_{jkl}
\bar C^{kl}_{\bar{i}}+ (1-\frac{\chi}{24})
G_{j \bar{i}}\,, \label{anom2}
\end{equation}
and special geometry, determine all correlation functions up
to holomorphic ambiguities. In Eq.~(\ref{anom2}), $\chi$ is the Euler
character of the manifold. A solution of the
recursion equations is given in terms of Feynman rules \cite{Bershadsky:1993cx}.

The propagators $S$, $S^i$, $S^{ij}$ for these Feynman rules are related
to the three point couplings $C_{ijk}$ as
\begin{equation}
\partial_{\bar{i}} S^{ij}= \bar{C}_{\bar{i}}^{ij}, \qquad
\partial_{\bar{i}} S^j = G_{i\bar{i}} S^{ij}, \qquad
\partial_{\bar{i}} S = G_{i \bar{i}} S^i.
\label{prop}
\end{equation}
By definition, the propagators $S$, $S^i$ and $S^{ij}$ are sections of
the bundles
$\mathcal{L}^{-2}\otimes \text{Sym}^m T$ with $m=0,1,2$.
The vertices of the Feynman rules are
given by the correlation functions $\mathcal{F}^{(g)}_{i_1\cdots
i_n}$.
The anomaly equation Eq.~(\ref{anom1}), as well as the definitions
in Eq.~(\ref{prop}), leave the freedom of adding holomorphic
functions under the $\ov{\partial}$ derivatives as integration
constants. This freedom is referred to as holomorphic ambiguities.


\subsection{Polynomial structure of higher genus amplitudes}\label{sec:polynomial}

In ref.\cite{Alim:2007qj} it was proven that the correlation functions
$\F{g}_{i_1\cdots i_n}$ are polynomials of degree $3g-3+n$ in the
generators $K_i,S^{ij},S^{i},S$ where a grading $1,1,2,3$ was assigned to these generators respectively. It was furthermore shown that by making a change of generators \cite{Alim:2007qj}
\begin{eqnarray}\label{shift}
\tilde{S}^{ij} &=& S^{ij}, \nonumber \\
\tilde{S}^i &=& S^i - S^{ij} K_j, \nonumber \\
\tilde{S} &=& S- S^i K_i + \frac{1}{2} S^{ij} K_i K_j, \nonumber\\
\tilde{K}_i&=& K_i\, , 
\end{eqnarray}
the $\mathcal{F}^{(g)}$ do not depend on $\tilde{K}_i$, i.e. $\partial \mathcal{F}^{(g)}/\partial \tilde{K}_i=0$. We will henceforth drop the tilde from the modified generators.

The proof relies on expressing the first non-vanishing correlation functions in terms of these generators. At genus zero these are
the holomorphic three-point couplings $\mathcal{F}^{(0)}_{ijk} = C_{ijk}$.
The holomorphic anomaly equation Eq.~(\ref{anom1}) can be integrated
using Eq.~(\ref{prop}) to
\begin{equation}
\mathcal{F}^{(1)}_i = \h C_{ijk} S^{jk} +(1-\frac{\chi}{24}) K_i
+ f_i^{(1)}, \label{sol2}
\end{equation}
with ambiguity $f_i^{(1)}$. As can be seen from
this expression, the non-holomorphicity of the correlation functions
only comes from the generators. Furthermore the special geometry relation (\ref{curvature}) can be integrated:
\begin{equation}
\Gamma^l_{ij} = \delta_i^l K_j + \delta^l_j K_i - C_{ijk} S^{kl} + s^l_{ij}\,,
\label{specgeom}
\end{equation}
where $s^l_{ij}$ denote holomorphic functions that are not fixed by the
special geometry relation, this can be used to derive the following equations which show the closure of the generators carrying the non-holomorphicity under taking derivatives \cite{Alim:2007qj}.\footnote{These equations are for the tilded generators of Equ. \ref{shift} and are obtained straightforwardly from the equations in ref.\cite{Alim:2007qj}}
\begin{eqnarray} \label{rel}
\partial_i S^{jk} &=& C_{imn} S^{mj} S^{nk} + \delta_i^j S^k +\delta_i^k S^j-s_{im}^j S^{mk} -s_{im}^k S^{mj} + h_i^{jk} \, , \nonumber\\
\partial_i S^j &=& C_{imn} S^{mj} S^n + 2 \delta_i^j S -s_{im}^j S^m -h_{ik} S^{kj} +h_i^j \, ,\nonumber\\
\partial_i S &=& \frac{1}{2} C_{imn} S^m S^n -h_{ij} S^j +h_i \, ,\nonumber\\
\partial_i K_j &=& K_i K_j -C_{ijn}S^{mn} K_m + s_{ij}^m K_m -C_{ijk} S^k + h_{ij} \, ,
\end{eqnarray}
where $h_i^{jk}, h^j_i$, $h_i$ and $h_{ij}$ denote
holomorphic functions. All these functions together with the functions
in Equ.(\ref{specgeom}) are not independent. It was shown in
ref. \cite{Alim:2008kp} (See also \cite{Hosono:2008ve}) that the
freedom of choosing the holomorphic functions in this ring reduces to holomorphic functions $\cE^{ij},\cE^j,\cE$ which can be added to the polynomial generators
\begin{eqnarray} \label{freedom}
\widehat{S}^{ij} &=& S^{ij} + \cE^{ij} \, ,\nonumber\\
\widehat{S}^{j} &=& S^{j} + \cE^{j} \, ,\nonumber\\
\widehat{S} &=& S + \cE \, .
\end{eqnarray}
All the holomorphic quantities change accordingly.

The topological string amplitudes now satisfy the holomorphic anomaly equations where the $\bar{\partial}_{\bar{i}}$ derivative is replaced by derivatives with respect to the polynomial generators \cite{Alim:2007qj}. 
\begin{eqnarray}
  \frac{\partial \F{g}_{i_1\dots i_n}}{\partial S^{ij}}-\frac{1}{2} \left( K_i \, \frac{\partial \F{g}_{i_1\dots i_n}}{\partial S^{j}}  +  K_j \, \frac{\partial \F{g}_{i_1\dots i_n}}{\partial S^{i}}\right) +\frac{1}{2} K_i K_j \frac{\partial \F{g}_{i_1\dots i_n}}{\partial S} = \nonumber\\
\frac{1}{2}  \sum_{r=0}^{g}\sum_{s=0}^{n} \frac{1}{s!(n-s)!} \sum_{\sigma \in S_n}
D_j\mathcal{F}^{(r)}_{i_{\sigma(1)} \dots i_\sigma(s)} D_k\mathcal{F}^{(g-r)}_{i_{\sigma(s+1)}\dots i_{\sigma(n)}} +\frac{1}{2} D_jD_k\mathcal{F}^{(g-1)}_{i_1\dots i_n} \label{inspol1}\\
\sum_i G_{i\bar{i}}  \frac{\partial \F{g}_{i_1\dots i_n}}{\partial K_{i}}=  -(2g-2+n-1)\sum_{s=1}^n G_{\bar{i}i_s} \mathcal{F}^{g}_{i_1\dots i_{s-1}i_s\dots i_n}\, . \label{inspol2}
\end{eqnarray}
This equation can be simplified by grouping powers of $K_i$
\cite{Hosono:2008ve}. 

\subsection{Constructing the generators} \label{generators}
The construction of the generators of the polynomial construction has been discussed in ref.\cite{Alim:2008kp}. The starting point is to pick a local coordinate $z_*$ on the moduli space such that $C_{*ij}$ is an invertible $n \times n$ matrix in order to rewrite Eq.(\ref{specgeom}) as
\begin{equation}\label{schoice}
S^{ij}=(C^{-1}_{*})^{ik} \left( \delta_*^j K_k +\delta_k^j K_{*} -\Gamma_{*k}^j +s_{*k}^j\right)
\end{equation}
The freedom in Eq.(\ref{freedom}) can be used to choose some of the $s_{ij}^k$ \cite{Alim:2008kp}. The other generators are then constructed using the equations (\ref{rel}) \cite{Alim:2008kp}:
\begin{eqnarray}\label{holchoice}
S^i &=& \frac{1}{2} \left( \partial_i S^{ii}  -C_{imn} S^{mi} S^{ni} + 2 s_{im}^i S^{mi} -h_i^{ii}\right) \,, \\
S&=& \frac{1}{2} \left(  \partial_i S^i -C_{imn} S^m S^{ni} +s_{im}^i S^m + h_{im} S^{mi} -h_i^i\right) \, .
\end{eqnarray}
In both equations the index $i$ is fixed, i.e. there is no summation over that index. The freedom in adding holomorphic functions to the generators of Eq.(\ref{freedom}) can again be used to make some choice for the functions $h^{ii}_i,h^i_i$, the other ones are fixed by this choice and can be computed from Eq.(\ref{rel}).

\subsection{Boundary conditions}

\subsubsection*{\it Genus 1}
The holomorphic anomaly equation at genus $1$  (\ref{anom2}) can be integrated to give:
\begin{equation}
\label{eq:27}
\mathcal{F}^{(1)}= \frac{1}{2} \left( 3+h^{2,1} -\frac{\chi}{12}\right) K +\frac{1}{2} \log \det G^{-1} +\sum_i s_i \log z_i + \sum_a r_a \log  \Delta_a \,,
\end{equation}
where $i=1,\dots,h^{2,1}$ and $a$ runs over the number of discriminant components. The coefficients $s_i$ and $r_a$ are fixed by the leading singular behavior of $\mathcal{F}^{(1)}$ which is given by \cite{Bershadsky:1993cx}
\begin{equation}
\mathcal{F}^{(1)} \sim -\frac{1}{24} \sum_i \log z_i \int_Z c_2 J_i \, ,
\end{equation}
for the algebraic coordinates $z_i$, for a discriminant $\Delta$ corresponding to a conifold singularity the leading behavior is given by 
\begin{equation} 
 \mathcal{F}^{(1)} \sim -\frac{1}{12} \log \Delta \,.
 \end{equation}

\subsubsection*{\it Higher genus boundary conditions}
The holomorphic ambiguity needed to reconstruct the full topological string amplitudes can be fixed by imposing various boundary conditions for $\F{g}$ at the boundary of the moduli space. As in Section~\ref{sec:periods-mirror-map} we assume that the boundary is described by simple normal crossing divisors $\overline{\cM} \setminus \cM = \bigcup_{i \in I} D_i$ for some finite set $I$.  

We can distinguish the various boundary conditions by looking at the monodromy $T_i$ of the Gauss--Manin connection $\nabla$ around a boundary divisor $D_i$. By the monodromy theorem~\cite{Landman:1973ab} we know that $T_i$ must satisfy 
\begin{equation}
  \left({T_i}^m - 1\right)^n = 0
  \label{eq:15}
\end{equation}
for $n\leq \dim Z^*+1$ and some positive integer $m$. The current understanding of the boundary conditions for $\F{g}$ seems to suggest that they can be classified according to the value of $n$. 

\subsubsection*{{\it The large complex structure limit}}

A point in the boundary is a large complex structure limit or a point of maximal unipotent monodromy if $n=\dim Z^*+1$ in~(\ref{eq:15}) and if $N_i = \log T_i$ satisfies certain conditions described in detail in~\cite{Cox:1999ab} and~\cite{Candelas:1994hw}.

The leading behavior of $\F{g}$ at this point (which is mirror to the large volume limit) was computed in \cite{Bershadsky:1993ta,Bershadsky:1993cx,Marino:1998pg,Gopakumar:1998ii,Faber:1998,Gopakumar:1998jq}. In particular the 
contribution from constant maps is 
\begin{equation} \label{constmaps}
 \mathcal{F}^{(g)}|_{q_a=0}= (-1)^g \frac{\chi}{2} \frac{|B_{2g} B_{2g-2}|}{2g\,(2g-2)\,(2g-2)!} \; , \quad g>1,
\end{equation}
where $q_a$ denote the exponentiated mirror map at this point.  

\subsubsection*{{\it Conifold-like loci}}

A divisor $D_i$ in the boundary is of conifold type if $n=2$ in~(\ref{eq:15}). If $m=1$ then $Z^*$ acquires a conifold singularity, if $m>1$ the singularity is not of conifold type but the physics behaves similarly. This singularity is often called a strong coupling singularity~\cite{Klemm:1996kv}. Singularities of both types appear at the vanishing of the discriminant $\Delta$. A well-known example for the case $m>1$ is the divisor given by the non-principal disciminant in the moduli space of the mirror of $\mP(1,1,2,2,6)[12]$ for which $m=2$. 

The leading singular behavior of the partition function $\F{g}$ at a conifold locus has been determined in \cite{Bershadsky:1993ta,Bershadsky:1993cx,Ghoshal:1995wm,Antoniadis:1995zn,Gopakumar:1998ii,Gopakumar:1998jq}
\begin{equation} \label{Gap}
 \mathcal{F}^{(g)}(t_c)= b \frac{B_{2g}}{2g (2g-2) t_c^{2g-2}} + O(t^0_c),
\qquad g>1
\end{equation}
Here $t_c\sim \Delta^{\frac{1}{m}}$ is the flat coordinate 
at the discriminant locus $\Delta=0$. For a conifold singularity $b=1$ and $m=1$. In particular the leading singularity in \eqref{Gap} as well as the absence of subleading singular terms follows from the Schwinger loop computation of \cite{Gopakumar:1998ii,Gopakumar:1998jq}, which computes the effect of the extra massless hypermultiplet 
in the space-time theory \cite{Vafa:1995ta}. The singular structure and the ``gap''  
of subleading singular terms have been also observed in the dual matrix model
\cite{Aganagic:2002wv} and were first used in \cite{Huang:2006si,Huang:2006hq} 
to fix the holomorphic ambiguity at higher genus. The space-time derivation of \cite{Gopakumar:1998ii,Gopakumar:1998jq} is
not restricted to the conifold case and applies also to the case $m>1$ 
singularities which give rise to a different spectrum of
extra massless vector and hypermultiplets in space-time. 
The coefficient of the Schwinger loop integral is a weighted trace over the spin of the particles~\cite{Vafa:1995ta, Antoniadis:1995zn} leading to the prediction $b=n_H-n_V$ for the coefficient of the leading singular term. The appearance of the prefactor $b$ in the case $m>1$ has been discussed in~\cite{Alim:2008kp} for the case of the local $\mF_2$ (see also~\cite{Haghighat:2009nr}).

\subsubsection*{{\it Orbifold loci}}
\label{sec:it-orbifold-loci}

A divisor $D_i$ in the boundary is of orbifold type if $n=1$ in~(\ref{eq:15}). In this case, the monodromy is of finite order. The leading singular behavior of the partition function $\F{g}$ at a such a divisor is expected to be regular~\cite{Bershadsky:1993cx}
\begin{equation}
  \label{eq:14}
  \F{g}(t_o) = O({t_o}^0), \qquad g > 1.
\end{equation}
where $t_o$ is the flat coordinate at the orbifold locus $D_i$. 

\subsubsection*{{\it The holomorphic ambiguity}}
\label{sec:it-holom-ambig}

\def\tz{\tilde{z}}
The singular behavior of $\F{g}$ is taken into account by the local ansatz
\begin{equation}\label{ansatzha}
\mathrm{hol. ambiguity}\sim \frac{p(\tz_i)}{\Delta^{(2g-2)}},
\end{equation}
for the holomorphic ambiguity near $\Delta=0$, 
where $p(\tz_i)$ is a priori a series in the local coordinates $\tz_i$ near the singularity. Patching together the local information at all the singularities with the boundary 
divisors with finite monodromy, it follows however that the numerator 
$p(z_i)$ is generically a polynomial of low degree in the $z_i$. Here $z_i$ denote the natural coordinates centered at 
large complex structure, $z_i=0\ \forall i$.
The finite number of coefficients in $p(z_i)$ is constrained by \eqref{Gap}. 


\section{Higher genus amplitudes for an elliptic fibration}\label{sec:highergenus}
In this section we use the polynomial construction together with the boundary conditions discussed previously to construct the higher genus topological string amplitudes for the example of the elliptic fibration which we discussed.

\subsection{Setup of the polynomials}
We start by setting up the polynomial construction as discussed in section \ref{sec:polynomial}. This involves using the freedom in choosing the generators in order to fix the holomorphic functions appearing in the derivative relations (\ref{rel}). We fix the choice of the polynomial generators such that these functions are rational expressions in terms of the coordinates in the large complex structure patch of the moduli space. For the holomorphic functions in the following we multiply lower indices by the corresponding coordinates and divide by the coordinates corresponding to upper indices. 
$$ A_i^j \rightarrow \frac{z_i}{z_j} A_i^j$$
With this convention we can express all the holomorphic functions appearing in the setup of the polynomial construction in terms of polynomials in the local coordinates. We start by fixing the choice of the generators $S^{ij}$ in Equs.(\ref{schoice},\ref{specgeom}):
\begin{align}
  s^1_{11} &= -2, & s^1_{12} &= -\frac{1}{3}, & s^1_{22} &= 0,\\
  s^2_{11} &= 0, & s^2_{12} &= 0, & s^2_{22} &= -\frac{4}{3}.
\end{align}

Then the following quantities are partly chosen by fixing the choice of the generators $S^i$ in Equ.(\ref{holchoice}) and the other quantities are then computed from Equs.(\ref{rel});
\begin{align}
  h^{11}_1 &=
  \frac{1}{9}-48\,z_{{1}}+\frac{5}{6}\,z_{{2}}-540\,z_{{1}}z_{{2}},\\
  h^{12}_1 &= -{\frac
    {5}{108}}-\frac{5}{4}\,z_{{2}}+20\,z_{{1}}+540\,z_{{1}}z_{{2}},\\
  h^{22}_1 &= -60\,z_{{1}}\left(1 - 27\,z_{{2}}\right),\\
  h^{11}_2 &= -60\,z_{{1}}z_{{2}},\\
  h^{12}_2 &= \frac{1}{9}+{\frac {5}{12}}\,z_{{2}}-48\,z_{{1}},\\
  h^{22}_2 &= -{\frac {23}{54}}+40\,z_{{1}}-\frac{5}{2}\,z_{{2}}-540\,z_{{1}}z_{{2}}.
\end{align}

We proceed by fixing the choice of the generator $S$ in Equ.(\ref{holchoice}) and compute from Equ.(\ref{rel})
\begin{align}
  h^1_1 &= {\frac {155}{27}}\,z_{{1}}-{\frac {25}{1296}}\,z_{{2}}+50\,z_{{1}}z_{{
2}},\\
  h^2_1 &= 0,\\
  h^1_2 &= -{\frac {5}{18}}\,z_{{2}}+120\,z_{{1}}z_{{2}},\\
  h^2_2 &= {\frac {155}{27}}\,z_{{1}} + {\frac {1055}{1296}}\,z_{{2}}+50\,z_{{1}}z_
{{2}}.
\end{align}
We further compute:
\begin{equation}
  h^1 = \frac {25}{23328},\quad  h^2 = -\frac {50}{3}\,z_{{1}}z_{{2}}.
\end{equation}
and
\begin{equation}
  h_{11} = \frac {5}{36},\quad  h_{12} = \frac {5}{108},\quad h_{22} = 0.
\end{equation}

With these choices the polynomial part of the higher genus amplitudes is entirely fixed by equations (\ref{inspol1}). However we need to supplement this polynomial part with the holomorphic ambiguities which are not captured by the holomorphic anomaly recursion and can be fixed by the boundary conditions discussed earlier. In order to implement the boundary conditions we make an ansatz for the ambiguities which will be discussed later. We then expand the polynomial part and the ansatz in the local special coordinates in the different patches of moduli space. In order to do this for the discussed example we first proceed by discussing the moduli space and its various loci.

\subsection{Moduli space and its compactification}
\label{sec:moduli-space-its}

To obtain a nice and useful description of the moduli space of complex
structures, we first need the secondary fan of the variety $W$. This is obtained from the columns of the Mori generators (\ref{chargevec}) which are (taking the primitive lattice vectors in $\mZ^2$)
\begin{align}
  \label{eq:17}
  b_1 &=(1,0), & b_2&=(0,1), & b_3 &= (1,-3), & b_4 &= (-1,0).
\end{align}
These vectors define the weighted projective space $\mP(1,1,3)$ blown
up in one point, with toric divisors $D_{(1,0)}$, $D_{(0,1)}$,
$D_{(1,-3)}$, $D_{(-1,0)}$, respectively. (The divisor $D_{(1,-3)}$ does not lie on the boundary of the moduli space~\cite{Candelas:1994hw} and will be neglected in the followoing.) This space is still singular, and we will discuss the
resolution of the singularities in the next subsection. 

We still have to remove the set where the hypersurface is singular,
i.e. the discriminant locus. This is also given in terms of the data
of toric geometry as follows: If $\theta$ is
any face of the polytope $\Delta^*$, we define $f_\theta(x) = \sum_{\nu_i
  \in \theta \cap \mZ^4} a_i \prod_i X^{\nu_{i}}$. The
polynomial is degenerate if for any face $\theta \subset \Delta^*$,
the system of polynomial equations
\begin{equation}
  \label{eq:19}
  f_\theta = X_1\pd{f}{X_1} = \dots = X_4 \pd{f}{X_4} = 0
\end{equation}
has no solution in the toric variety. This yields that the discriminant locus is given by the divisors
\begin{equation}
  \label{eq:18}
  D_1 = \{ \Delta_1 = 0\},\qquad D_2 = \{ \Delta_2 = 0 \}.
\end{equation}
with $\Delta_1$ and $\Delta_2$ given in~(\ref{eq:Discriminant}).

In the following we will use the following abbreviations
\begin{equation}
  \label{eq:12}
  \begin{aligned}
    \zbar_1 &= 432 z_1, & \zbar_2 &= -27z_2
  \end{aligned}
\end{equation}

These divisors intersect each other as follows. From $\Delta_1 =
(1-\zbar_1)^3-{\zbar_1}^2\zbar_2$, we see that there is a tangency of order 3
between $D_{(0,1)}$ and $D_1$ at the point $(1,0)$.
Writing $\Delta_1 = (1 - 3\zbar_1 + 3{\zbar_1}^2) + {\zbar_1}^3\Delta_2$ we
see that there is a triple intersection of $D_1$ and $D_2$ intersect transversally
in the two points $(\zbar_1,\zbar_2) = \left(\zbar_{\pm},
1\right)$ with $\zbar_\pm=\frac{1}{2}\left(1\pm i \frac{\sqrt{3}}{3}\right)$. By changing to the variables to $w_1= \frac{1}{\zbar_1}$ we write $\Delta_1 = 
-w_1(3 - 3w_1 + {w_1}^2)+\Delta_2$ and we have a triple intersection
of $D_1$, $D_2$, and $D_{(-1,0)}$ in $(w_1,\zbar_2) = (0,1)$.

\subsubsection*{\it Resolution of singularities}
\label{sec:resol-sing}

We want a compactification of the complex structure moduli space by
divisors with normal crossings. To achieve this we must resolve the
singularities of $\mP(1,1,3)$ and resolve all nonnormal crossings of
$D_1$ and $D_2$ with any of the other divisors. Moreover, we will need
a set of local coordinates near of each normal crossing.

The singularities of $\mP(1,1,3)$ can be taken care of by toric
geometry. Resolving them amounts to subdividing the secondary fan and
this introduces three further generators $b_5=(1,-1)$, $b_6=(1,-2)$
and $b_7=(0,-1)$, and the corresponding toric divisors $D_{(1,-1)}$,
$D_{(1,-2)}$ and $D_{(0,-1)}$. Toric geometry also provides us with
the local coordinates near each intersection point of the toric
divisors. They are determined by the generators of the cone dual to
the maximal cone spanned by the corresponding generators. E.g. the
dual cone to $\langle 0, b_5, b_6 \rangle$ is spanned by the vectors $(2,1)$ and
$(-1,-1)$, hence the corresponding local coordinates are $\left({\zbar_1}^2\zbar_2,
\frac{1}{\zbar_1\zbar_2}\right)$. A summary can be found in Table~\ref{tab:local_coords}.

In order to obtain normal crossings with $D_1$ and $D_2$ we first
consider the resolution of the singularity of the hypersurface
$W=x^3-y^4=0$ in $(0,0)$. We view the hypersurface $W=0$ as a divisor
$D$ in $\mC^2$. The resolution can be performed in terms of four blow-ups.

At the first blow-up, we introduce a $\mP^1$ with homogeneous coordinates $(u_0:v_0)$ such that
$u_0x-v_0y=0$. We denote this exceptional divisor by $E_0$. In the
coordinate patch $u_0=1$ we have $x=v_0y$ and the singularity becomes
$W=y^3({v_0}^3-y)$. $W=0$ now consists of the
components $E_0=\{y=0\}$ and $D=\{{v_0}^3-y=0\}$ which do not
intersect transversely in $(v_0,y) = (0,0)$. On the other hand, in the
coordinate patch $v_0=1$, we have $y=u_0x$ and the singularity becomes
$W=x^3(1-{u_0}^4x)$.  $W=0$ consists of the
components $E_0=\{x=0\}$ and $D=\{1-{u_0}^4x=0\}$ which do not
intersect at all. Hence, we focus on the patch ${u_0}=1$ with local
coordinates $(v_0, y)$ and resolve further. 

At the second blow-up, we introduce a $\mP^1$ with homogeneous coordinates $(u_1,v_1)$ such that
$u_1v_0-v_1y=0$. We denote this exceptional divisor by $E_1$. In the
coordinate patch $u_1=1$ we have $v_0=v_1y$ and the singularity
becomes $W=y^4({v_1}^3y^2-1)$. $W=0$ now consists of
the components $E_1=\{y=0\}$ and $D=\{{v_1}^2y^2-1=0\}$ which do not
intersect. On the other hand, in the coordinate patch $v_1=1$, we have
$y=u_1v_0$ and the singularity becomes
$W={u_1}^3{v_0}^4({v_0}^2-u_2)$.  $W=0$
consists of the components $E_1=\{v_0=0\}$, $E_0=\{u_1=0\}$ and
$D=\{{v_0}^2-u_1=0\}$ which do not intersect transversely in
$(v_0,u_1)=(0,0)$. Hence, we focus on the patch $v_1=1$ with
local coordinates $(v_0,u_1)$ and resolve further. 

At the third blow-up, we introduce a $\mP^1$ with homogeneous coordinates $(u_2:v_2)$ such that
$u_2v_0-v_2u_1=0$. We denote this exceptional divisor by $E_2$. In the
coordinate patch $u_2=1$ we have $v_0=v_2u_1$ and the singularity
becomes
$W={u_1}^6{v_2}^2(u_1{v_2}^2-1)$. $W=0$
consists of the components $E_2=\{u_1=0\}$, $E_1=\{v_2=0\}$ and
$D=\{u_1{v_2}^2-1=0\}$ which do not intersect. On the other hand, in
the coordinate patch $v_2=1$, we have $u_1=u_2v_0$ and the singularity
becomes $W={u_2}^3{v_0}^6(v_0-u_2)$.
$W=0$ consists of the components $E_2=\{v_0=0\}$, $E_0=\{u_2=0\}$ and
$D=\{v_0-u_2=0\}$ which do not intersect transversely in
$(v_0,u_2)=(0,0)$. Hence, we focus on the patch $v_2=1$ with local
coordinates $(v_0,u_2)$ and resolve further.  

At the fourth and final blow-up, we introduce a $\mP^1$ with homogeneous coordinates $(u_3:v_3)$ such that $u_3v_0-v_3u_0=0$. We denote this exceptional divisor by $E_3$. In the coordinate patch $u_3=1$ we have $v_0=v_3u_2$ and the singularity becomes $W={u_2}^{10}{v_3}^6(v_3-1)$. $W=0$ consists of the components $E_3=\{u_2=0\}$, $E_2=\{v_3=0\}$, $D=\{v_3-1=0\}$ which do not intersect. On the other hand, in the coordinate patch $v_3=1$, we have $u_4=u_3v_0$ and the singularity becomes $W={u_3}^3{v_0}^{10}(1-u_3)$.  $W=0$ consists of the components $E_3=\{v_0=0\}$, $E_0=\{u_3=0\}$ and $D=\{1-u_3=0\}$ which do intersect transversely in $(u_3,v_0)=(0,0)$.  Hence, we have completely resolved the singularity.

We see that $E_0 \cap E_3 = \{v_0=0,u_3=0\}$, $E_3 \cap D =
\{v_0=0,u_3=1\} = \{u_2=0,v_3=1\}$ and $E_0 \cap D =
\emptyset$. Moreover, in the other patch, $E_3 \cap E_2 =
\{u_2=0,v_3=0\}$, $E_2 \cap D =\emptyset$, and $E_0 \cap E_2 =
\emptyset$. Since $E_1$ does not appear anymore, $E_3 \cap E_1 =
\emptyset$, its intersections can only be seen in the previous patch
with coordinates $(u_1,v_2)$ and are $E_0 \cap E_1 = \emptyset$ and
$E_1 \cap E_2 = \{u_1=0,v_2=0\}$.

Now, we apply this to the divisors in  the moduli space of the mirror
of $\mP(1,1,1,6,9)[18]$. After the first blow-up $W={v_0}^3 - y$
describes a tangency of order 3 which locally can be identified with
the tangency of $D_1$ and $D_{(0,1)}$. This yields
$D=D_1$, $E_0 = D_{(0,1)}$ with local coordinates
\[v_0 = 1 - \zbar_1, \qquad y  = -{\zbar_1}^3\zbar_2.\] From this we get
\begin{equation}
  \label{eq:297}
  \begin{aligned}
    u_1 &= \frac{y}{v_0} = -\frac{{\zbar_1}^3\zbar_2}{1-\zbar_1}, & v_1  &=
    \frac{v_0}{y} = -\frac {1-\zbar_1}{{\zbar_1}^3\zbar_2},\\
    u_2 &= \frac{u_1}{v_0} = -\frac{{\zbar_1}^3\zbar_2}{(1-\zbar_1)^2}, & v_2 &=
    \frac{v_0}{u_1} = -\frac {(1-\zbar_1)^2}{{\zbar_1}^3\zbar_2},\\
    u_3 &= \frac{u_2}{v_0} = -\frac{{\zbar_1}^3\zbar_2}{(1-\zbar_1)^3}, & v_3 &=
    \frac{v_0}{u_2} = -\frac {(1-\zbar_1)^3}{{\zbar_1}^3\zbar_2}.
 \end{aligned}
\end{equation}
With these identifications we find for the local coordinates near the four
intersections of these divisors
\begin{equation}
  \label{eq:298}
  \begin{aligned}
    D_1 \cap E_3: & \left( 1 +
      \frac{{\zbar_1}^3\zbar_2}{\left(1-\zbar_1\right)^3}, 1-\zbar_1 \right)\\
    D_{(0,1)} \cap E_3: & \left( -
      \frac{{\zbar_1}^3\zbar_2}{\left(1-\zbar_1\right)^3}, 1-\zbar_1 \right)\\
    E_2 \cap E_3: & \left( -
      \frac{{\zbar_1}^3\zbar_2}{\left(1-\zbar_1\right)^2}, -\frac{(1-\zbar_1)^3}{{\zbar_1}^3\zbar_2} \right)\\
    E_1 \cap E_2: & \left( -
      \frac{{\zbar_1}^3\zbar_2}{1-\zbar_1}, -\frac{(1-\zbar_1)^2}{{\zbar_1}^3\zbar_2} \right)
 \end{aligned}
\end{equation}

Similarly, the triple intersection $W=u_2v_0(v_0-u_2)$ after the third
blowup locally can be identified with the triple intersection of $D_1$, $D_2$, and
$D_{(-1,0)}$. For this purpose, 
we set \[u_2 = 1-\zbar_2, \qquad v_0 = \alpha\, w_1\] where $\alpha =
{w_1}^2-3w_1+3$ which is nonzero at $w_1=0$.  This yields $D = D_1$,
$E_0 = D_2$ and $E_2=D_{(-1,0)}$. From this we get (recalling
$w_1=\frac{1}{\zbar_1}$ neglecting factors of $\alpha$) 
\begin{equation}
  \label{eq:21}
  \begin{aligned}
    u_3 &= \frac{u_2}{v_0} = \zbar_1(1-\zbar_2), & v_3 &=
    \frac{v_0}{u_2} = \frac {1}{\zbar_1(1-\zbar_2)}.    
  \end{aligned}
\end{equation}
Relabeling the exceptional divisor $E_3$ by $E_0$ we find for the
local coordinates near the two intersection points
\begin{equation}
  \label{eq:299}
  \begin{aligned}
    D_1 \cap E_0: & \left(
      \frac{1}{{\zbar_1}^2}\left((1-\zbar_1)^3+{\zbar_1}^3\zbar_2 \right), \frac{1}{\zbar_1} \right)\\
    D_2 \cap E_0: & \left( \zbar_1(1-\zbar_2), \frac{1}{\zbar_1} \right)\\
    D_{(0,1)} \cap E_0: & \left( \frac{1}{\zbar_1(1-\zbar_2)} , 1-\zbar_2\right)
 \end{aligned}
\end{equation}
This concludes the construction of the compactification of the moduli space with normal crossing divisors. We summarize the local coordinates in Table~\ref{tab:local_coords}.
\begin{table}[htb]
  \centering
  $
  \begin{array}{|c|c|c|}
    \hline
    \textrm{Crossing} & 
    \textrm{Local coordinates}\\
    \hline
    D_{(1,0)} \cap D_{(0,1)} & \left (\zbar_1,\zbar_2\right) \\
    D_{(1,0)} \cap D_{(1,-1)} & \left (\zbar_1\zbar_2,{\zbar_2}^{-1}\right) \\
    D_{(1,0)} \cap D_2 & \left
      (\zbar_1,1-\zbar_2\right) \\
    D_{(1,-2)} \cap D_{(1,-1)} & \left((\zbar_1\zbar_2)^{-1},{\zbar_1}^2{\zbar_2}\right)\\
    D_2 \cap E_0 & \left({\zbar_1}(1-\zbar_2),
  \frac{1}{{\zbar_1}}\right)\\
    D_{(-1,0)} \cap D_{(0,-1)} & \left ({\zbar_1}^{-1},{\zbar_2}^{-1}\right) \\
    D_{(-1,0)} \cap D_{(0,1)} & \left ({\zbar_1}^{-1},\zbar_2\right) \\
    D_{(-1,0)} \cap E_0 & \left(\frac{1}{{\zbar_1}(1-\zbar_2)}, 1-\zbar_2\right)\\
    \left(D_1 \cap D_2\right)_+ &
    \left(1-\frac{\zbar_1}{\zbar_+},\frac{1-\zbar_2}{1-\frac{\zbar_1}{\zbar_+}}\right)\\
    \left(D_1 \cap D_2\right)_- &
    \left(1-\frac{\zbar_1}{\zbar_-},\frac{1-\zbar_2}{1-\frac{\zbar_1}{\zbar_-}}\right)\\
    D_1 \cap E_0 & \left(\frac{(1-\zbar_1)^3+{\zbar_1}^3\zbar_2}{{\zbar_1}^2},\frac{1}{{\zbar_1}}\right)\\
    E_3 \cap E_2 & \left(
      -\frac{{\zbar_1}^3\zbar_2}{(1-\zbar_1)^2}, -\frac{(1-\zbar_1)^3}{{\zbar_1}^3\zbar_2} \right)\\
    E_3 \cap D_{(0,1)} & \left(1-\zbar_1, -\tfrac{{\zbar_1}^3\zbar_2}{(1-\zbar_1)^3} \right)\\
    E_3 \cap D_1 & \left( 1-\zbar_1,
      1+\tfrac{{\zbar_1}^3\zbar_2}{(1-\zbar_1)^3}\right)\\
    E_1 \cap E_2 & \left( -
      \frac{{\zbar_1}^3\zbar_2}{1-\zbar_1}, -\frac{(1-\zbar_1)^2}{{\zbar_1}^3\zbar_2} \right)\\
   \hline
  \end{array}
  $
  \caption{}
  \label{tab:local_coords}
\end{table}
where $\zbar_\pm = \frac{1}{2} \pm i \frac{\sqrt{3}}{6}$. 
\begin{figure}[htb!]
\centering%
\includegraphics[width=120mm]{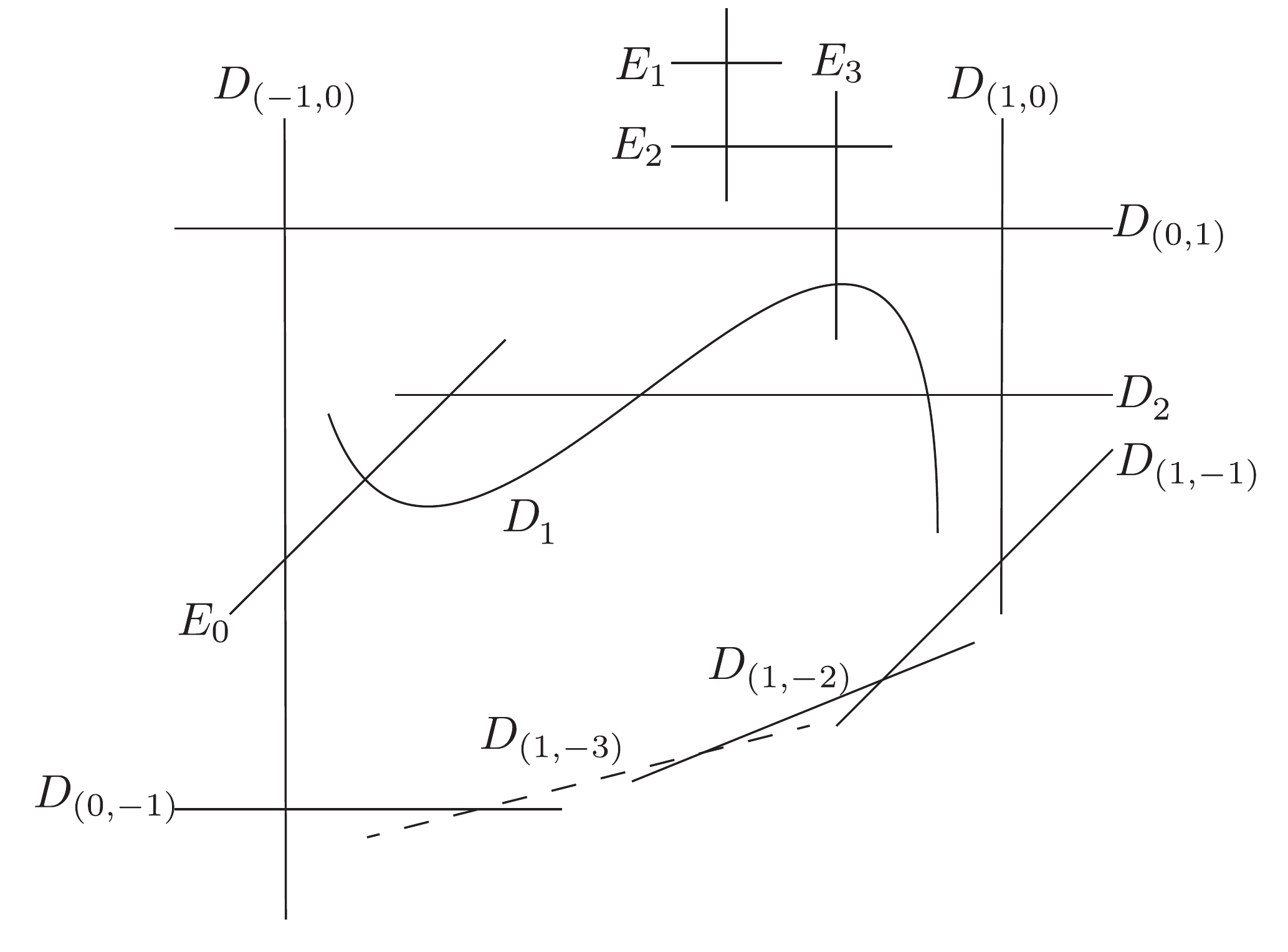}
\caption{The blown--up moduli space}
\label{fig:P113}
\end{figure}
We give a sketch for the compactification in Figure~\ref{fig:P113}. The divisor $D_{(1,-3)}$ is drawn with a dashed line since in it not in the boundary of the moduli space. Under the action of the symmetry $I$ given in~\eqref{eq:25}, we have
\begin{equation}
  \label{eq:13}
  \begin{aligned}
    I(D_{(1,-2)}) &= E_1, & I(D_{(1,-1)}) &= E_2, & I(D_{(1,0)}) &= E_3, & I(D_1) &= D_2, \\
    I(D_{(0,1)}) &= D_{(0,1)}, & I(D_{(0,-1)}) &= D_{(0,-1)}, &
    I(D_{(-1,0)}) &= D_{(-1,0)}, & I(E_0) &= E_0.
  \end{aligned}
\end{equation}
For a sketch of the compactification in coordinates in which this symmetry becomes manifest, see~\cite{Candelas:1994hw}.

\subsection{Periods and flat coordinates at the boundary points}
\label{sec:periods-mirror-maps}

Consider the intersection points $p$ of the boundary divisors listed in Table~\ref{tab:local_coords}. We again denote the local coordinates near one of these points $p$ by $y$. For each of the first nine intersections $p$ of (the remaining ones can be obtained by applying the symmetry $I$) we determine the Gauss--Manin connection. This can be done in two ways, starting from the results at the large complex structure point reviewed in Section~\ref{sec:mirrorsymmetry}. Either one performs the change of variables from $z$ to $y$ given in this table in the Picard--Fuchs equation~(\ref{PF}) and then reads off the connection matrix as discussed in Appendix~\ref{sec:gaussmanin}, or one transforms the connection matrix using the gauge transformation law for this change of variables. In both cases, one needs to specify a basis of periods near the intersection of interest. We choose it to be the same everywhere and as in~(\ref{filtration}) and express it in terms of differential operators acting on a period as
\begin{equation}
  \label{eq:20}
  1, \quad \theta_1, \quad \theta_2, \quad {\theta_1}^2, \quad {\theta_1\theta_2}, \quad  {\theta_1}^2\theta_2.
\end{equation}
where $\theta_i = y_i\pd{}{y_i}$.

Once we have the connection matrices $A_i(y)$, we can determine their residues. The residues are then used in two ways. First, they allow us to compute the index of the monodromy about the divisors intersecting $p$. Second, they enter into the solutions of the Picard--Fuchs equations as discussed in Section~\ref{sec:mirrorsymmetry}. For the residues we find (the residues for $D_{(1,0)}$ and $D_{(0,1)}$ have been displayed in~(\ref{eq:Residue}) )
\begin{equation}
 \begin{aligned}
   \label{eq:26}
    \Res_{D_{(1,-1)}} &\sim \left( \begin {array}{cccccc} 0&1&0&0&0&0\\ \noalign{\medskip}0&0&1&0
&0&0\\ \noalign{\medskip}0&0&0&1&0&0\\ \noalign{\medskip}0&0&0&0&0&0
\\ \noalign{\medskip}0&0&0&0&\frac{1}{3}&0\\
\noalign{\medskip}0&0&0&0&0&\frac{2}{3}
\end {array} \right) &
   \Res_{D_{2}} & \sim \left( \begin {array}{cccccc} 0&0&0&0&0&0\\ \noalign{\medskip}0&2&0&0
&0&0\\ \noalign{\medskip}0&0&1&1&0&0\\ \noalign{\medskip}0&0&0&1&0&0
\\ \noalign{\medskip}0&0&0&0&0&0\\ \noalign{\medskip}0&0&0&0&0&1
\end {array} \right) 
  \end{aligned}
\end{equation}
\begin{equation*}
  \begin{aligned}
    \Res_{D_{(1,-2)}} &\sim \left( \begin {array}{cccccc} 0&1&0&0&0&0\\ \noalign{\medskip}0&0&1&0
&0&0\\ \noalign{\medskip}0&0&0&1&0&0\\ \noalign{\medskip}0&0&0&0&0&0
\\ \noalign{\medskip}0&0&0&0&\frac{2}{3}&0\\
\noalign{\medskip}0&0&0&0&0&\frac{1}{3}
\end {array} \right) &
  \Res_{E_{0}} &\sim \left( \begin {array}{cccccc} 1&0&0&0&0&0\\ \noalign{\medskip}0&\frac{1}{6}&0
&0&0&0\\ \noalign{\medskip}0&0&\frac{5}{6}&0&0&0\\ \noalign{\medskip}0&0&0&\frac{7}{6}
&0&0\\ \noalign{\medskip}0&0&0&0&{\frac {11}{6}}&0
\\ \noalign{\medskip}0&0&0&0&0&1\end {array} \right)
  \end{aligned}
\end{equation*}
\begin{equation*}
  \begin{aligned}
  \Res_{D_{(-1,0)}} &\sim \left( \begin {array}{cccccc} \frac{1}{6}&0&0&0&0&0\\ \noalign{\medskip}0&\frac{5}{6}
&0&0&0&0\\ \noalign{\medskip}0&0&\frac{1}{6}&0&0&0\\ \noalign{\medskip}0&0&0&\frac{5}{6}&0&0\\ \noalign{\medskip}0&0&0&0&\frac{1}{6}&0\\ \noalign{\medskip}0&0&0&0&0
&\frac{5}{6}\end {array} \right) &
   \Res_{D_{(0,-1)}} &\sim \left( \begin {array}{cccccc} \frac{1}{18}&0&0&0&0&0\\ \noalign{\medskip}0&{
\frac {5}{18}}&0&0&0&0\\ \noalign{\medskip}0&0&{\frac {7}{18}}&0&0&0
\\ \noalign{\medskip}0&0&0&{\frac {11}{18}}&0&0\\ \noalign{\medskip}0&0
&0&0&{\frac {13}{18}}&0\\ \noalign{\medskip}0&0&0&0&0&{\frac {17}{18}}
\end {array} \right)
  \end{aligned}
\end{equation*}
\begin{equation*}
  \begin{aligned}
       \Res_{D_{1}} &\sim \left( \begin {array}{cccccc} 0&0&0&0&0&0\\ 
\noalign{\medskip}0&2&0&0
&0&0\\ \noalign{\medskip}0&0&1&1&0&0\\ \noalign{\medskip}0&0&0&1&0&0
\\ \noalign{\medskip}0&0&0&0&0&0\\ \noalign{\medskip}0&0&0&0&0&1
\end {array} \right) &
    \Res_{E_3} &\sim \left( \begin {array}{cccccc} 0&1&0&0&0&0\\
        \noalign{\medskip}0&0&1&0 &0&0\\
        \noalign{\medskip}0&0&0&1&0&0\\ \noalign{\medskip}0&0&0&0&0&0
        \\ \noalign{\medskip}0&0&0&0&0&0\\
        \noalign{\medskip}0&0&0&0&0&0
      \end {array} \right)
  \end{aligned}
\end{equation*}
\begin{equation*}
  \begin{aligned}
   \Res_{E_2} &\sim \left( \begin {array}{cccccc} 0&1&0&0&0&0\\ \noalign{\medskip}0&0&1&0
&0&0\\ \noalign{\medskip}0&0&0&1&0&0\\ \noalign{\medskip}0&0&0&0&0&0
\\ \noalign{\medskip}0&0&0&0&\frac{1}{3}&0\\
\noalign{\medskip}0&0&0&0&0&\frac{2}{3}
\end {array} \right) &
   \Res_{E_1} &\sim \left( \begin {array}{cccccc} 0&1&0&0&0&0\\ \noalign{\medskip}0&0&1&0
&0&0\\ \noalign{\medskip}0&0&0&1&0&0\\ \noalign{\medskip}0&0&0&0&0&0
\\ \noalign{\medskip}0&0&0&0&\frac{2}{3}&0\\
\noalign{\medskip}0&0&0&0&0&\frac{1}{3}
\end {array} \right)\\
  \end{aligned}
\end{equation*}
We note that the monodromy matrices $\Res_{D_1}$, $\Res_{D_2}$ appear at the various intersection points always with an eigenvalue $1$ of multiplicity $3$, but the multiplicities of the eigenvalues $0$ and $2$ are different at different points of
intersection. This does not matter here, and can easily be remedied by multiplying the basis elements~(\ref{eq:20}) with appropriate powers of $y_i$. We have summarized the behaviour of the various monodromy matrices in Table~\ref{tab:Monodromies}. (This has first been obtained in~\cite{Candelas:1994hw}. The monodromies about $D_{(1,-1)}$ and $D_{(1,-2)}$ can be related to the one about $D_{(1,0)}$ through the local toric geometry~\cite{Berglund:1993pw}.)
\begin{table}[htb!]
  \centering
  $
  \begin{array}{|c|c|}
    \hline
    D_{(1,0)} & (T-1)^4 = 0\\
    D_{(0,1)} & (T-1)^3 = 0\\
    D_{(1,-1)} & (T^3-1)^4 = 0\\
    D_{(1,-2)} & (T^3-1)^4 = 0\\
    D_{(0,-1)} & T^{18}-1 = 0\\
    D_{(-1,0)} & T^6-1 = 0\\
    D_{1} & (T-1)^2=0\\
    D_2 & (T-1)^2 = 0\\
    E_0 & T^6-1 = 0\\
    E_1 & (T^3-1)^4 = 0\\
    E_2 & (T^3-1)^4 = 0\\
    E_3 & (T-1)^4 = 0\\
   \hline
  \end{array}
  $
  \caption{}
  \label{tab:Monodromies}
\end{table}
We note here that by~\cite{Candelas:1994hw} the generators of the monodromy group $\Gamma$ are $D_{(0,-1)}$ and $D_1$. The generators of the monodromy subgroup $\Gamma_{\text{ell}}$ corresponding to the elliptic fiber are $D_{(1,0)}$ and ${D_{(0,-1)}}^3$.

For the solutions of the Picard--Fuchs equations we only give an example, for the other points the results are analogous. The local coordinates at the intersection $D_{(1,0)} \cap D_{(1,-1)}$ read
\begin{equation}
  \label{eq:197}
  \begin{aligned}
    y_{{1}}&=-11664\,z_1z_2,&y_{{2}}&=-\frac{1}{27\,z_2}
  \end{aligned}
\end{equation}
The residue matrices at $y_i = 0$ have been displayed in~(\ref{eq:26}). The solutions of the Picard--Fuchs operators take the form
\begin{equation}
  \label{eq:199}
  \begin{aligned}
    \pi_0(y) &= s_0(y)\\
    \pi_1(y) &= s_0\log\left(y_1{y_2}^{\frac{2}{3}}\right) + s_1(y)\\
    \pi_2(y) &= s_0\log\left(y_1{y_2}^{\frac{2}{3}}\right)^2 + 2\,s_1(y)\log\left(y_1{y_2}^{\frac{2}{3}}\right)+s_2(y)\\
    \pi_3(y) &=  s_0\log\left(y_1{y_2}^{\frac{2}{3}}\right)^3 + 3\,s_1\log\left(y_1{y_2}^{\frac{2}{3}}\right)^2 + 3\,s_2(y)\log\left(y_1{y_2}^{\frac{2}{3}}\right)+s_3(y)\\
    \pi_4(y) &= {y_2}^{\frac{1}{3}}s_4(y)\\
    \pi_5(y) &= {y_2}^{\frac{2}{3}}s_5(y)\\
  \end{aligned}
\end{equation}
with
\begin{equation}
  \label{eq:200}
  \begin{aligned}
     s_0(y) &= 1+{\frac {5}{36}}\,y_{{1}}y_{{2}}{h}^{2}+O \left( {h}^{4} \right)\\
     s_1(y) &= {\frac {31}{36}}\,y_{{1}}y_{{2}}{h}^{2}+O \left( {h}^{4} \right)\\
     s_2(y) &={\frac {5}{18}}\,y_{{1}}y_{{2}}{h}^{2}+O \left( {h}^{4} \right)\\
     s_3(y) &=-y_{{2}}h+ \left( -{\frac {9}{40}}\,{y_{{2}}}^{2}-\frac{5}{6}\,y_{{1}}y_{{2}}
 \right) {h}^{2}+O \left( {h}^{3} \right) \\
     s_4(y) &=1+\frac{1}{24}\,y_{{2}}h+ \left( {\frac {4}{315}}\,{y_{{2}}}^{2}+{\frac {5}{
72}}\,y_{{1}}y_{{2}} \right) {h}^{2}+O \left( {h}^{3} \right) \\
     s_5(y) &=1+ \left( -{\frac {5}{18}}\,y_{{1}}+\frac{2}{15}\,y_{{2}} \right) h+O \left( 
{h}^{2} \right) 
 \end{aligned}
\end{equation}
We obtain the symplectic form $Q$ at $p$ in the same way as the connection matrices $A_i$, by changing the variables in~(\ref{eq:Q}). Then, inserting the solutions $\pi_i(y)$ yields the intersection form
\begin{equation}
  \label{eq:201}
  Q = \left( \begin {array}{cccccc} 0&0&0&\frac{1}{9}&0&0\\ \noalign{\medskip}0&0&-
\frac{1}{27}&0&0&0\\ \noalign{\medskip}0&\frac{1}{27}&0&0&0&0\\ \noalign{\medskip}-\frac{1}{9}
&0&0&0&0&0\\ \noalign{\medskip}0&0&0&0&0&\frac{1}{27}\\ \noalign{\medskip}0&0&0
&0&-\frac{1}{27}&0\end {array} \right)
\end{equation}
This allows us the choose the flat coordinates as follows:
\begin{equation}
  \label{eq:202}
  \begin{aligned}
    t_1(y) &= \frac{\pi_1(y)}{\pi_0(y)} = \log\left(y_1{y_2}^{\frac{2}{3}}\right)+{
\frac {31}{36}}\,y_{{1}}y_{{2}}{h}^{2}+O \left( {h}^{4} \right) \\
   t_2(y) &= \frac{\pi_4(y)}{\pi_0(y)} = {y_2}^{\frac{1}{3}}\left(1+\frac{1}{24}\,y_{{2}}h+O \left( {h}^{2} \right)\right)
  \end{aligned}
\end{equation}

\subsubsection*{{\it The partition function for $g=2,3$}}
\label{sec:part-funct-g=2}

Having the flat coordinates at all the intersections points at the
boundary at hand, we can proceed to apply the boundary conditions
discussed in Section~\ref{sec:recursion}. In genus 1, we use $c_2 J_1
= 102$ and $c_2 J_2 = 36$ to fix the $s_i$ in~\eqref{eq:27} to be
$s_1=-\frac{15}{4}$, $s_2=-\frac{7}{6}$, and furthermore we find $r_1=r_2 = -\frac{1}{6}$.

From Table~\ref{tab:Monodromies} we see that $D_{(0,-1)}$, $D_{(-1,0)}$, and $E_0$ are of orbifold type, while $D_1$ and $D_2$ are of conifold type. 

The condition that $\F{g}$ be regular at a divisor with finite monodromy, i.e. at $D_{(0,-1)}$, and $D_{(-1,0)}$ ensures that the holomorphic function $p^{(g)}(z)$ is a polynomial. The degrees $(d_1,d_2)$ of its monomials are bounded by
\begin{align}
  \label{eq:3}
  d_1 &\leq 7(g-1), & d_2 &\leq 6(g-1)-1, & 3d_2 - d_1 &\leq 9(g-1).
\end{align}
In addition, regularity at $D_{(1,-1)}$ fixes the coefficients of the monomials with degrees $3d_2-d_1 > 3(2g-2)$, while regularity at $D_{(-1,0)}$ fixes those with $d_1 > 6(g-1)$. The divisor $E_0$ does not yield additional conditions.

Since $D_1$ and $D_2$ are of conifold type, we can use the expansion~(\ref{Gap}). In order to do so, we have to take into account that the flat coordinates $t_1, t_2$ obtained from the above process are only determined up to normalization. Hence we expect relations $t_i = k_i t_{c,i}$, $i=1,2$, where $t_{c,i}$ are the flat coordinates in the expansion~(\ref{Gap}). The gap condition from this expansion yields an overdetermined systems of relations among the remaining coefficients of $p^{(g)}(z)$. This system has a unique solution depending only on the parameter $k_1$. This normalization factor could in principle be determined by an explicit analytic continuation of the periods $\pi(z)$ at the large complex structure limit to the periods $\pi(y)$ at $D_1 \cap D_2$, though this is highly complicated. 

Finally, at the large complex structure limit we can apply the Gopakumar--Vafa expansion~\cite{Gopakumar:1998jq}:
\begin{equation*}
  \cF(Z,t,\lambda)  = \frac{c(t)}{\lambda^2} + l(t) + \sum_\beta \sum_{m > 0} \sum_{r\geq 0} \frac{1}{m} n^{(g)}_\beta(Z) \left(2\sin\left(\tfrac{m\lambda}{2}\right)\right)^{2g-2} q^{m\beta}
\end{equation*}
where $c(t)$ and $l(t)$ are a cubic and linear polynomials, respectively, depending on topological invariants of $Z$. Using the fact that there are no degree $1$ curves of genus $2$ in the base, $n^{(2)}_{0,1} = 0$ allows us to determine $k_1$ as well. The Gopakumar--Vafa invariants $n^{(g)}_\beta$ are listed in Appendix~\ref{sec:gv-invariants}. The resulting expressions for the ambiguities $f^{(2)}(z)$ and $f^{(3)}(z)$ can be found in Appendix~\ref{sec:holom-ambig}. For $g>3$ the computations turn out to be too involved. Moreover, we expect that the boundary conditions known so far, will not be sufficient to fix the holomorphic ambiguity. We observe that the $\cF^{(g)}$ also show a particular behaviour at the other boundary divisors $D_i$, however, it is not possible to formulate it just from the resulting expression.

\subsection{Recursion in terms of modular forms of $\tS\tL(2,\mZ)$}
Having computed the topological string partition function up to genus $3$ we proceed in the following with exploring the manifestation of the $\tS\tL(2,\mZ)$ subgroup of the modular group. To do so we examine the large complex structure expansion of $\F{g}$ in terms of the special coordinates. We need further to choose a section of the corresponding line bundle $\mathcal{L}^{2-2g}$. We do so by fixing the gauge $\pi_0(z)=1$, where $\pi_0$ is the analytic solution at large complex structure given in Equ.(\ref{eq:22}). The special, flat coordinates in this patch of moduli space are given by
\begin{equation}
t_E:=t_1=\frac{\pi_1}{\pi_0}\, ,\quad t_B:=t_2=\frac{\pi_2}{\pi_0}\, , \quad q_E:=q_1=e^{2\pi i t_1}\,, \quad q_B:=q_2=e^{2\pi i t_2}.
\end{equation}
where the periods $\pi_i$ are given in the Appendix~\ref{sec:gaussmanin}.
We consider the functions
\begin{equation}
F^{(g)}(t_E,t_B)= \pi_0(z(t))^{2g-2}\, \F{g}(z(t))\, ,
\end{equation}
and expand these in the exponentiated base modulus $q_B$:
\begin{equation}
F^{(g)}(t_E,t_B)= \sum_{n} f^{(g)}_n(t_E) {q_B}^n=\sum_n \frac{1}{n!} \frac{\partial^n F^{(g)}}{\partial {q_B}^n}\Big{|}_{q_B=0} {q_B}^n\, ,
\end{equation}
we find that $f^{(g)}_n$ can be written as 
$$f^{(g)}_n=P^{(g)}_n (E_2,E_4,E_6) \frac{q_E^{3n/2}}{\Delta^{3n/2}}\, ,$$
where $P^{(g)}_n$ denotes a quasi-modular form constructed out of the Eisenstein series $E_2,E_4,E_6$ of weight $2g+18n-2$, some examples of these are given in appendix (\ref{appendix:expansion})  we furthermore find that the $f^g_n$ satisfy the following recursion:
\begin{equation} \label{refinedrecursion}
 \frac{\partial f^{(g)}_n}{\partial E_2}=-\frac{1}{24} \sum_{h=0}^g \sum_{s=1}^{n-1} s (n-s) f^{(h)}_s f^{(g-h)}_{n-s} +\frac{n(3-n)}{24} f^{(g-1)}_n \, .
\end{equation}
This recursion is analogous to a recursion which was conjectured for higher genus in refs. \cite{Hosono:1999qc,Hosono:2002xj}. The geometry considered in these works was that of a $\frac{1}{2}K_3$. The recursion at genus $0$ was motivated by a recursion in the BPS state counting of the non-critical string \cite{Minahan:1997ct,Minahan:1997if,Minahan:1998vr} and its relation to the prepotential of the geometry used to construct these \cite{Klemm:1996hh}.\footnote{More recently this geometry has also been studied in \cite{Sakai:2011xg}.} 

The recursion at genus zero can be verified explicitly either from the construction of the polynomial expressions from integrals of the underlying Seiberg-Witten type curve \cite{Minahan:1997ct,Minahan:1997if} or from the properties of the Picard-Fuchs equations \cite{Hosono:1999qc}. The higher genus version of the equation is verified for low genera by the explicit construction of the polynomials.
In particular, the explicit knowledge of the holomorphic ambiguities $f^{(2)}$ and $f^{(3)}$ allow us to determine the $E_2$ independent part of the polynomials $P^{(g)}_n$ which is not determined by~\eqref{refinedrecursion}. 
Moreover, the higher genus version is conjectured to be equivalent to the BCOV anomaly equation \cite{Hosono:2002xj,Hosono:2008ve}. In the following we want to relate qualitatively the equation (\ref{refinedrecursion}) to the anomaly equations for the amplitudes with insertions in its polynomial form (\ref{inspol1},\ref{inspol2}).

We work with the coordinates centered at large complex structure $z_1$ and $z_2$ and consider the free energy with $n$ insertions w.r.t $z_2$:
$$F^{(g)}_n:=(\pi_0)^{2g-2} \mathcal{F}^{(g)}_{\underbrace{\textrm{\tiny{2\dots 2}}}_{n}}$$
since $z_2$ is not the flat coordinate, the insertions are defined using covariant derivatives on $T^*\mathcal{M}$. We will use however that $z_2=q_2+\dots$ and hence to leading order derivatives w.r.t $q_2$ are captured by the amplitudes with insertions w.r.t $z_2$. 

We are now interested in the appearance of $E_2$ in the $q_2\rightarrow 0$ limit in the polynomial generators of the full problem, we find an occurrence in two of the generators:\footnote{Since $S^{22}$ is a section of $\mathcal{L}^{-2}$ we fix a section by multiplying by $\pi_0^2$, we moreover have $\pi_0|_{q_2=0}=E_4^{1/4}$.}
\begin{eqnarray}
\left(\frac{S^{22}}{z_2^2}\, \right)|_{q_2=0}&=&-\frac{1}{12} E_2 +E_4^{1/2} +\frac{1}{12} \frac{E_6}{E_4}\, ,\\
K_1|_{q_2=0}&=&\frac{E_4^{3/2}}{\Delta} \left(E_2 E_4-E_6 \right)\, .
\end{eqnarray}
We hence have
\begin{equation}
\frac{\partial F^{(g)}_n}{\partial E_2} \Big{|}_{q_2=0}= \left(\frac{\partial F^{(g)}_n}{\partial S^{22}} \,\frac{\partial S^{22}}{\partial E_2}+\frac{\partial F^{(g)}_n}{\partial K_1}\, \frac{\partial  K_1}{\partial E_2}\right)\Big|_{q_2=0}\, ,
\end{equation}
the two terms on the r.h.s can be computed from Equs.(\ref{inspol1},\ref{inspol2}). The second of which vanishes in this case due to the vanishing of the K\"ahler metric $G_{\bar{1}2}$ on the r.h.s of Equ.(\ref{inspol2}) in the limit $q_2\rightarrow 0$.

We therefore have from (\ref{inspol1}):
\begin{equation}
\frac{\partial F^{(g)}_{n}}{\partial S^{22}}= \frac{1}{2}  \sum_{h=0}^{g}\sum_{s=0}^{n} 
D_2{F}^{(h)}_{s} D_2{F}^{(g-h)}_{n-s} +\frac{1}{2} D_2D_2{F}^{(g-1)}_{n} 
\end{equation}
and furthermore:
\begin{equation}\label{refrec2}
\frac{\partial F^{(g)}_n}{\partial E_2} \Big{|}_{q_2=0} = -\frac{z_2^2}{24} \left( \sum_{h=0}^{g}\sum_{s=0}^{n} 
D_2{F}^{(h)}_{s} D_2{F}^{(g-h)}_{n-s} + D_2D_2{F}^{(g-1)}_{n} \right)\Big{|}_{q_2=0}
\end{equation}
We further compute $z_2 \Gamma_{22}^2|_{q_2=0}=-1$ and note that 
$$z_2 D_2 F^{(g)}_n|_{q_2=0} =\left(\theta_2 F^{(g)}_{n}-n z_2 \Gamma_{22}^2 F^{(g)}_n\right)|_{q_2=0} = n \left(F^{(g)}_n\right)|_{q_2=0}\, ,$$
Relating the $f^{(g)}_n \sim F^{(g)}_{n}\big|_{q_2=0}$ it is possible to see the characteristic traits of equation (\ref{refinedrecursion}). Due to the multiplication with $z_2^2$ the non-vanishing contribution of the first term on the r.h.s of (\ref{refrec2}) is coming from the product of the connections with prefactor $s(n-s)$, from the second term, a contribution of $n(n+1)$ is coming from the contribution of the product of the two connections. Further contributions come from derivatives acting on the connections. This completes our qualitative relation of refined recursion (\ref{refinedrecursion}) to the polynomial form of the holomorphic anomaly equation with insertions (\ref{inspol1}). A more thorough matching of the two equations is beyond the scope of this work and will be discussed elsewhere.

\subsection{Further examples}
\label{sec:further-examples}

The expansion~\eqref{refinedrecursion} also holds for other elliptic fibrations. We present here some more examples. The first is a section of the anti-canonical bundle over the resolved weighted projective space $\mP(1,1,1,3,6)$. The charge vectors for this geometry are given by:
\begin{equation}
  \label{chargevec2}
\begin{array}{cccccccc}
&x_0&x_1&x_2&x_3&x_4&x_5&x_6\\
(l^1)=&(-4&2&1&1&0&0&0\, )\, ,\\
(l^2)=&(\hskip6pt 0&0&0&-3&1&1&1\, )\, .\\
\end{array}
\end{equation}
If we take the derivative with respect to $E_2(2\tau)$ instead of $E_2(\tau)$, then~\eqref{refinedrecursion} holds with the first initial condition given as
\begin{equation}
  \label{eq:314}
  f^{(0)}_1 = \frac{3}{8}\,F_{{2}}{G_{{2}}}^{3} \left( 16\,{F_{{2}}}^{4}-51\,{F_{{2}}}^{2}{
G_{{2}}}^{2}+51\,{G_{{2}}}^{4} \right) {\Delta} ^{-3/2},
\end{equation}
where $F_2$ and $G_2$ are modular forms of weight $2$ and generate the
ring of modular forms for $\Gamma(2)$. They can be expressed in terms of Jacobi theta functions as
\begin{equation}
  \label{eq:323}
  \begin{aligned}
    F_2(\tau) &= \theta_2(\tau)^4 + \theta_3(\tau)^4,\\
    G_2(\tau) &= \theta_2(\tau)^4 - \theta_3(\tau)^4.
  \end{aligned}
\end{equation}
The same is true, if we consider a section of the anti-canonical bundle over the resolved weighted projective space $\mP(1,1,1,3,3)$ whose charge vectors are
\begin{equation}
  \label{chargevec3}
\begin{array}{cccccccc}
&x_0&x_1&x_2&x_3&x_4&x_5&x_6\\
(l^1)=&(-3&1&1&1&0&0&0\, )\, ,\\
(l^2)=&(\hskip6pt 0&0&0&-3&1&1&1\, )\, .\\
\end{array}
\end{equation}
Taking the derivative with respect to $E_2(3\tau)$ instead of $E_2(\tau)$, then~\eqref{refinedrecursion} holds with initial condition 
\begin{equation}
 \label{eq:319}
  f^{(0)}_1 = 9\,E_{{1}} \left( {E_{{1}}}^{6}-87\,F_{{3}}{E_{{1}}}^{3}+2349\,{F_{{3}
}}^{2} \right)  \left( {E_{{1}}}^{3}-27\,F_{{3}} \right) ^{3} \Delta^{-3/2} ,
\end{equation}
where $E_1$ and $F_3$ are modular forms of weight $1$ and $3$,
respectively, and generate the
ring of modular forms for $\Gamma_1(3)$. They can be expressed in terms
of the Dedekind eta functions as
\begin{equation}
  \label{eq:322}
  \begin{aligned}
    E_1(\tau) &= \frac{\left(\eta(\tau)^{12} + 27
        \eta(3\tau)^{12}\right)^{\frac{1}{3}}}{\eta(\tau)\eta(3\tau)},\\
    F_3(\tau) &= \frac{\eta(3\tau)^9}{\eta(\tau)^3}.
  \end{aligned}
\end{equation}
Another elliptic fibration whose associated congruence subgroup is $\Gamma_1(3)$ is the degree $(3,3)$ hypersurface in $\mP^2 \times \mP^2$. Its charge vectors are
\begin{equation}
  \label{chargevec4}
\begin{array}{cccccccc}
&x_0&x_1&x_2&x_3&x_4&x_5&x_6\\
(l^1)=&(-3&1&1&1&0&0&0\, )\, ,\\
(l^2)=&(\hskip6pt -3&0&0&0&1&1&1\, )\, .\\
\end{array}
\end{equation}
and the first initial condition for the recursion is
\begin{equation}
  \label{eq:321}
  F^{(0)}_1=27\,E_1\, \left( 7\,{E_1}^{3}+54\,F_3
 \right) \Delta^{-1/2}.
\end{equation}
A similar example as~\eqref{chargevec2} and~\eqref{chargevec3} is a
complete intersection of two sections of the anti-canonical bundle over the resolved weighted projective space $\mP(1,1,1,3,3,3)$ whose charge vectors are
\begin{equation}
  \label{chargevec8}
\begin{array}{cccccccccc}
&x_{0,1}&x_{0,2}&x_1&x_2&x_3&x_4&x_5&x_6&x_7\\
(l^1)=&(-2&-2&1&1&1&1&0&0&0\, )\, ,\\
(l^2)=&(\hskip6pt 0&\phantom{-}0&0&0&0&-3&1&1&1\, )\, .\\
\end{array}
\end{equation}
Taking the derivative with respect to $E_2(4\tau)$ instead of $E_2(\tau)$, then~\eqref{refinedrecursion} holds with initial condition 
\begin{equation}
 \label{eq:30}
  f^{(0)}_1 = 3 {E_1}^3 {F_1}^9 \left( 4{E_1}^4-13 {E_1}^2{F_1}^2+13{F_1}^4\right) \Delta^{-3/2} ,
\end{equation}
where $E_1$ and $F_1$ are modular forms of weight $1$, and generate the
ring of modular forms for $\Gamma_1(4)$. They can be expressed in terms
of the Dedekind eta functions as
\begin{equation}
  \label{eq:31}
  \begin{aligned}
    E_1(\tau) &= \frac{\left(\eta(\tau)^{8} + 16
        \eta(4\tau)^{8}\right)^{\frac{1}{2}}}{\eta(2\tau)^2},\\
    F_1(\tau) &= \frac{\eta(\tau)^4}{\eta(2\tau)^2}.
  \end{aligned}
\end{equation}

The argument of the previous subsection also applies to elliptic
fibrations over Hirzebruch surfaces $\mF_n$, $n=0,1,2$. They have more than one base modulus. For example, the elliptic fibration given by the charge vectors
\begin{equation}
  \label{chargevec5}
\begin{array}{ccccccccc}
&x_0&x_1&x_2&x_3&x_4&x_5&x_6&x_7\\
(l^1)=&(-6&3&2&1&0&0&0&0\, )\, ,\\
(l^2)=&(\hskip6pt 0&0&0&-2&1&1&0&0\, )\, ,\\
(l^3)=&(\hskip6pt 0&0&0&-2&0&0&1&1\, )\, .\\
\end{array}
\end{equation}
has base $\mF_0$. In this case the recursion~\eqref{refinedrecursion} takes the following form
\begin{equation}
  \label{eq:312}
  \begin{aligned}
    \pd{f^{(g)}_{m,n}}{E_2} =&
    -\frac{1}{24}\left(2mn-2m-2n\right)f^{(g-1)}_{m,n}\\
    &-\frac{1}{24}\sum_{h=0}^g\sum_{s=0}^m\sum_{t=0}^n
    \left(s(n-t) + t(m-s)\right) f^{(g-h)}_{s,t}f^{(h)}_{m-s,n-t}
  \end{aligned}
\end{equation}
with first initial condition
\begin{equation}
  \label{eq:324}
  f^{(0)}_{0,1} = -2\frac{E_4E_6}{\Delta}.
\end{equation}
and $f^{(g)}_{m,n} = f^{(g)}_{n,m}$. 
The fact that the $f^{(g)}_{m,n}$ can be expressed in the form 
$f^{(g)}_{m,n}=P^{(g)}_{m,n} (E_2,E_4,E_6) {\Delta^{-m-n}}$ where $P^{(g)}_{m,n} (E_2,E_4,E_6)$ is a quasi--modular form of weight $2g-2+12m+12n$ has already been observed in~\cite{Klemm:2004km}. 

Next, we consider an elliptic fibration over the Hirzebruch surface
$\mF_1$ which has two phases. In the phase with charge vectors 
\begin{equation}
  \label{chargevec6}
\begin{array}{ccccccccc}
&x_0&x_1&x_2&x_3&x_4&x_5&x_6&x_7\\
(l^1)=&(-6&3&2&1&0&0&0&0\, )\, ,\\
(l^2)=&(\hskip6pt 0&0&0&-2&1&1&0&0\, )\, ,\\
(l^3)=&(\hskip6pt 0&0&0&-1&0&-1&1&1\, )\, .\\
\end{array}
\end{equation}
the recursion turns out to be
\begin{equation}
  \label{eq:312}
  \begin{aligned}
    \pd{f^{(g)}_{m,n}}{E_2} =&
    -\frac{1}{24}\left(2mn-2m-n-n^2\right)f^{(g-1)}_{m,n}\\
    &+\frac{1}{24}\sum_{h=0}^g\sum_{s=0}^m\sum_{t=0}^n
    \left(t(n-t) -s(n-t) - t(m-s)\right) f^{(g-h)}_{s,t}f^{(h)}_{m-s,n-t}
  \end{aligned}
\end{equation}
with first initial conditions
\begin{equation}
  \label{eq:324}
  \begin{aligned}
    f^{(0)}_{0,1} &= \frac{E_4}{\Delta^{1/2}}, & f^{(0)}_{0,1} &= -2\frac{E_4E_6}{\Delta}\,.
  \end{aligned}
\end{equation}
In this case, the quasi-modular form $P^{(g)}_{m,n} (E_2,E_4,E_6)$ has weight $2g-2+12m+6n$. The modularity of $f^{(0)}_{0,1}$ has been analyzed in detail in~\cite{Klemm:1996hh}.

Finally, for the elliptic fibration over $\mF_2$ given by the charge vectors
charge vectors 
\begin{equation}
  \label{chargevec7}
\begin{array}{ccccccccc}
&x_0&x_1&x_2&x_3&x_4&x_5&x_6&x_7\\
(l^1)=&(-6&3&2&1&0&0&0&0\, )\, ,\\
(l^2)=&(\hskip6pt 0&0&0&-2&1&1&0&0\, )\, ,\\
(l^3)=&(\hskip6pt 0&0&0&0&0&-2&1&1\, )\, .\\
\end{array}
\end{equation}
we find that the recursion turns out to be
\begin{equation}
  \label{eq:28}
  \begin{aligned}
    \pd{f^{(g)}_{m,n}}{E_2} =&
    -\frac{1}{24}\left(2mn-2m-2n^2\right)f^{(g-1)}_{m,n}\\
    &+\frac{1}{24}\sum_{h=0}^g\sum_{s=0}^m\sum_{t=0}^n
    \left(2t(n-t) -s(n-t) - t(m-s)\right) f^{(g-h)}_{s,t}f^{(h)}_{m-s,n-t}
  \end{aligned}
\end{equation}
with first initial conditions
\begin{equation}
  \label{eq:324}
  \begin{aligned}
    f^{(0)}_{1,0} &= -2\frac{E_4E_6}{\Delta}, & f^{(0)}_{0,1} & =0.
  \end{aligned}
\end{equation}
In this case, the quasi-modular form $P^{(g)}_{m,n} (E_2,E_4,E_6)$ has
weight $2g-2+12m$.

As last example, we consider Schoen's Calabi-Yau, i.e. a complete
intersection of two equations of degrees $(3,1,0)$ and $(0,1,3)$, respectively, in
$\mP^2\times\mP^1\times\mP^2$, i.e. the charge vectors are 
\begin{equation}
  \label{chargevec9}
\begin{array}{ccccccccccc}
&x_{0,1}&x_{0,2}&x_1&x_2&x_3&x_4&x_5&x_6&x_7&x_8\\
(l^1)=&(-3&0&1&1&1&0&0&0&0&0\, )\, ,\\
(l^2)=&(-1&-1&0&0&0&1&1&0&0&0\, )\, ,\\
(l^3)=&(\hskip6pt 0&-3&0&0&0&0&0&1&1&1\, )\, .\\
\end{array}
\end{equation}
This is an elliptic fibration over the rational elliptic surface
$\text{dP}_9$ studied in detail in~\cite{Hosono:1997hp} (see
also~\cite{Braun:2007vy}). For simplicity, we have restricted the
K\"ahler classes of the rational elliptic surface to the class of the
fiber and the section. The recursion turns out to be 
\begin{equation}
  \label{eq:32}
  \begin{aligned}
    \pd{f^{(g)}_{m,n}}{E_2} =&
    -\frac{1}{24}\left(9mn+3n^2-3n\right)f^{(g-1)}_{m,n}\\
    &+\frac{1}{24}\sum_{h=0}^g\sum_{s=0}^m\sum_{t=0}^n
    \left(-s(n-t) - t(m-s)\right) f^{(g-h)}_{s,t}f^{(h)}_{m-s,n-t}
  \end{aligned}
\end{equation}
with first initial conditions
\begin{equation}
  \label{eq:33}
  \begin{aligned}
    f^{(0)}_{1,0} &= 81\frac{1}{\Delta^{1/6}}, & f^{(0)}_{0,1} & =0.
  \end{aligned}
\end{equation}
In this case, the quasi-modular form $P^{(g)}_{m,n} (E_2,E_1,F_1)$ for $\Gamma_1(3)$
has weight $2g-2+2m$ with $E_1$ and $F_1$ given in~\eqref{eq:322}. The
modularity of $f^{(0)}_{1,0}$ has been proven in~\cite{Zagier:1998ab}.

\section{Conclusions}
\label{conclusion}

In this work we studied topological string theory  and mirror symmetry on an elliptically fibered CY.  We computed higher genus amplitudes for this geometry using their polynomial structure and appropriate boundary conditions.  The implementation of the boundary conditions required the use of techniques to single out the preferred coordinates on the deformation space of complex structures on the B-model side of topological strings. To do this we used the Gauss-Manin connection and the special, flat coordinates which could be found in various loci in the moduli space. At the large volume limiting point on the A-side which is mirror to the B-model large complex structure limit, the topological string free energies reduce to the Gromov-Witten generating functions allowing us thus to make predictions for these invariants at genus 2 and 3 in their resummed version giving the GV integer BPS degeneracies.

Having computed the higher genus topological string amplitudes we showed that these carry an additional interesting structure which exhibits the elliptic fibration. Namely the order by order expansion in terms of the moduli of the base of the elliptic fibration can be expressed in terms of the characteristic modular forms of $\tS\tL(2,\mZ)$ which is a subgroup of the full modular group due to the elliptic fibration. Along with this refined expansion in terms of $E_2,E_4$ and $E_6$ we found a refined anomaly equation which could be related to the holomorphic anomaly equations of BCOV for the correlation functions. This type of anomaly is the analog of an anomaly which was studied in the study of BPS states of exceptional non-critical strings \cite{Minahan:1997ct,Minahan:1997if,Minahan:1998vr} which are captured by the prepotential of the geometry used in their construction \cite{Klemm:1996hh}. It was furthermore shown in \cite{Minahan:1998vr} that this anomaly is related to an anomaly found in the study of partition functions of $\mathcal{N}=4$ topological SYM theory \cite{Vafa:1994tf}. The anomaly for the that latter theory on $\mP^2$ found in \cite{Vafa:1994tf} marks the first physical appearance of what became to be know as mock modular forms (See ref.\cite{Zagier:2007} for an introduction). The relation of the non-holomorphicity of mock modular forms and the recursion at genus $0$ was further studied in \cite{Manschot:2010xp,Manschot:2010nc,Alim:2010cf,Manschot:2011ym}. The recursion found in this work (\ref{refinedrecursion}) is expected to shed more light on the higher rank $\mathcal{N}=4$ topological SYM theory on $\mP^2$, since the main example of this paper is an elliptic fibration over $\mP^2$ and the elliptic fibration structure is the analogous setup to ref.\cite{Minahan:1998vr}. It would be furthermore interesting to give the higher genus amplitudes an interpretation in the SYM theory.

\subsection*{Acknowledgments}
We would like to thank S. Hosono, D. van Straten, K. Wendland and Timm
Wrase for helpful discussions. We would furthermore like to thank A. Klemm, J. Manschot and T. Wotschke for coordinating the submission of their work. M.A. would like to thank Michael Hecht, Dominique L\"ange and Peter Mayr for discussions and collaborations on related projects, where the recursion discussed in this paper was first observed. M.A. is supported by the DFG fellowship AL 1407/1-1.


\appendix

\section{Gauss-Manin connection matrices}
\label{sec:gaussmanin}
The vector $w(z)$ with $2h^{2,1}+2$ components:
\begin{equation}\label{filtrationApp}
w(z)= \left(\Omega(z)\, \quad \theta_1\Omega(z),\theta_2 \Omega(z)\quad \theta_1^2 \Omega(z),\theta_1\theta_2 \Omega(z)\, ,\quad \theta_1^2 \theta_2 \Omega(z)\, \right)^t\, .
\end{equation}
was picked such that its entries span the filtration quotient groups $(F^3,F^2/F^3,F^1/F^2,F^0/F^1)$ of respective orders $(1,h^{2,1},h^{2,1},1)$. Further multiderivatives of $\Omega(z)$ can be expressed in terms of the elements of this vector using the Picard-Fuchs equations, derivatives and linear combinations thereof. We find the following relation for the remaining double derivative:
\begin{equation}
  \label{eq:Degree2relation}
   {\theta_{{1}}}^{2} ={\frac {3\left(\theta_{{2}}\theta_{{1}}+144\,z_{{1}}
\theta_{{1}}+20\,z_{{1}}\right)}{\Delta_3}} .
\end{equation}
as well as relations for the triple derivatives, for example:

\begin{equation}
  \label{eq:Degree3relation}
  \begin{aligned}
    {\theta_{{1}}}^{3}&={\frac
      {3\left(164\,z_{{1}}\theta_{{1}}+53568\,{z_{{1}}
          }^{2}\theta_{{1}}+20\,z_{{1}}+1296\,\theta_{{2}}\theta_{{1}}z_{{1}}+
          8640\,{z_{{1}}}^{2}+3\,{\theta_{{2}}}^{2}\theta_{{1}}+60\,\theta_{{2}}
          z_{{1}}\right)}{ {\Delta_3} ^{2}}},\\
    {\theta_{{1}}}^{2}\theta_{{2}}&={\frac {3\,\theta_{{2}} \left( 20\,z_{
{1}}+144\,z_{{1}}\theta_{{1}}+\theta_{{1}}\theta_{{2}} \right) }{\Delta_3}}
  \end{aligned}
\end{equation}

The fourth order derivatives can be expressed in terms of the Gauss-Manin connection acting on the period matrix:
\begin{equation}
  \left(\theta_i- A_i(z) \right) \Pi(z)_{\beta}^{\phantom{\beta} \alpha} =0\, ,\quad i=1,\dots,h^{2,1}\, ,
\end{equation}

In the following we give these matrices at the large complex complex structure limit for the example discussed in this work:
\begin{equation}
  \label{eq:As}
  \begin{aligned}
    A_1(z) &=
\left( \begin {array}{cccccc} 0&1&0&0&0&0\\ \noalign{\medskip}{
\frac {60\, z_{{1}}}{\Delta_3}}&{\frac {432\, z_{{1}}}{\Delta_3}}&0&\frac{3}{\Delta_3}&0&0
\\ \noalign{\medskip}0&0&0&1&0&0\\ \noalign{\medskip}0&0&{\frac {
60\, z_{{1}}}{\Delta_3}}&{\frac {432\, z_{{1}}}{\Delta_3}}&0&
\frac{3}{\Delta_3}\\ \noalign{\medskip}0&0&0&0&0
&1\\ \noalign{\medskip}{\frac {3\,a_1 }{ \Delta_3\, \Delta_1 }}&{\frac {3\,a_2 }{
    \Delta_3\, \Delta_1 }}&{\frac {3\,a_3 }{ \Delta_3\, \Delta_1 }}&{\frac {3\,a_4 }{ \Delta_3\,
    \Delta_1 }}&{\frac { 60\,z_{{1}} {\Delta_3} ^{2}\Delta_2 }{ \Delta_1}}&\frac
  {a_5 }{ \Delta_3\, \Delta_1 }
\end {array} \right), \\
    A_2(z) &=
\left( \begin {array}{cccccc} 0&0&1&0&0&0\\ \noalign{\medskip}0&0&0&1
&0&0\\ \noalign{\medskip}0&0&0&0&1&0\\ \noalign{\medskip}0&0&0&0&0&1
\\ \noalign{\medskip} \frac{a_6}{{\Delta_3}^2\,\Delta_2} &\frac{a_7}{{\Delta_3}^2\,\Delta_2} &\frac{a_8}{{\Delta_3}^2\,\Delta_2} &\frac{a_9}{{\Delta_3}^2\,\Delta_2}&{\frac {-27\, z_{{2}}}{\Delta_2}}&\frac{a_9}{{\Delta_3}^2\,\Delta_2}
\\ \noalign{\medskip} \frac{a_1}{\Delta_1} &\frac{a_2}{\Delta_1} &\frac{a_3}{\Delta_1} 
&\frac{a_3}{\Delta_1}&\frac{60\,z_1 a_9}{\Delta_1}&\frac{a_{10}}{\Delta_1}\end {array} \right).
  \end{aligned}
\end{equation}
with
\begin{equation}
  \label{eq:as}
  \begin{aligned}
    a_1 &= 720\,{z_{{1}}}^{2}z_{{2}} \left(5 +  91152
\,z_{{1}}\right) ,\\
    a_2 &= -12\, z_{{2}}z_{{1}} \left( 5-
12960\,z_{{1}}-35645184\,{z_{{1}}}^{2} \right) ,\\
    a_3 &= 180
\, z_{{2}}z_{{1}} \left( 1-2160\,z_{{1}}+1679616\,{z_{{1}}}^{2}
 \right), \\
    a_4 &= -36\, z_{{2}}z_{{1}} \left( 5-8640\,z_{{1}} -71103744\,{z_{{1}}
}^{2}\right), \\
    a_5 &= 432\, z_{{1}} \left(\Delta_1(z) +30233088\,{z_{{1}}}^{
2}z_{{2}} \right), \\
    a_6 &= -120\,z_1z_2\left(1-864\,z_1\right),\\
    a_7 &= -2\,z_{{2}} \left( 1-1266\,z_
{{1}}  + 546912\,{z_{{1}}}^{2}\right), \\
    a_8 &= -6\, z_{{2}} \left( 1-804\,z_{{1}}+147744\,{z_{
{1}}}^{2} \right), \\
    a_9 &= 9\, z_{{2}}\left( 1-1296\,z_{{1}}+559872\,{z_{{1}}
}^{2} \right),\\
    a_{10} &= 4353564672\, {{z_{{1}}}^{3}z_{{2}}}.
  \end{aligned}
\end{equation}
A fundamental solution is given by
\begin{equation}
  \label{eq:22}
  \begin{aligned}
    \pi_0(z)  &= s_0(z),\\
    \pi_1(z)  &= s_0(z) \log z_1 + s_1(z),\\
    \pi_2(z)  &= s_0(z) \log z_2 + s_2(z),\\
    \pi_3(z)  &= s_0(z) \left(\frac{9}{2} \left(\log z_1\right)^2 + 3\, \log z_1 \log z_2\right)  + s_1(z) \log z_2 + s_2(z) \log z_1 + s_3(z),\\
    \pi_4(z)  &= s_0(z) \left( \frac{9}{2}\, \left( \log  z_{{1}}
 \right) ^{2}+3\,\log  z_{{1}} \log  z_{{
2}} +\frac{1}{2}\, \left( \log  z_{{1}} \right) ^{2}
 \right) \\ 
&\phantom{=}+ s_{{2}}(z) \left(3\,\log  z_{{1}}
+\log  z_{{2}} \right) + s_{{4}}(z),\\
    \pi_5(z)  &= s_0(z) \left( \frac{3}{2}\, \left(\log z_1\right)^3 + \frac{3}{2}\, \left(\log z_1\right)^2\log z_2 + \frac{1}{2}\, \log z_1\left(\log z_2\right)^2\right)\\
   &\phantom{=} + \frac{s_1(z) }{2}\left(\log z_2\right)^2 +s_2(z) \left(\frac{3}{2} \left(\log z_1\right)^2 + \log z_1 \log z_2\right) \\
   &\phantom{=}+ s_3(z) \log z_2 + s_4(z) \log z_1 + s_5(z),
  \end{aligned}
\end{equation}
where
\begin{equation}
  \label{eq:23}
  \begin{aligned}
    s_0(z) &= 1+{\frac {5}{36}}\,z_{{1}}+{\frac {385}{5184}}\,{z_{{1}}}^{2} + O(z^3),\\
    s_1(z) &= {\frac {13}{18}}\,z_{{1}}-{\frac {2}{27}}\,z_{{2}}+{\frac {719}{1728}}
\,{z_{{1}}}^{2}-{\frac {5}{243}}\,{z_{{2}}}^{2}+{\frac {5}{972}}\,z_{{
2}}z_{{1}} + O(z^3),\\
    s_2(z) &={\frac {5}{12}}\,z_{{1}}+\frac{2}{9}\,z_{{2}}+{\frac {385}{1152}}\,{z_{{1}}}^{
2}+{\frac {5}{81}}\,{z_{{2}}}^{2}-{\frac {5}{324}}\,z_{{2}}z_{{1}}
 + O(z^3),\\
    s_3(z) &= -\frac{1}{3}\,z_{{2}}+{\frac {13}{4}}\,{z_{{1}}}^{2}-{\frac {47}{324}}\,{z_{{2
}}}^{2} + O(z^3),\\
    s_4(z) &= {\frac {15}{4}}\,z_{{1}}+{\frac {10183}{768}}\,{z_{{1}}}^{2} + O(z^3),\\
    s_5(z) &= -\frac{15}{2}\,z_{{1}}+\frac{2}{3}\,z_{{2}}-{\frac {965}{256}}\,{z_{{1}}}^{2}+{\frac {13}{108}}\,{z_{{2}}}^{2}-{\frac {5}{108}}\,z_{{2}}z_{{1}}+ O(z^3).
  \end{aligned}
\end{equation}

\section{Holomorphic ambiguity}
\label{sec:holom-ambig}

\begin{equation}
  \label{eq:316}
  \begin{aligned}
    f^{(2)}(z)=&{\frac {1}{155520}}\, \left( -111885\,\zbar_{{1}}+25523\,\zbar_{{2}}+671310\,{
        \zbar_{{1}}}^{2}+111447\,\zbar_{{2}}\zbar_{{1}}-56842\,{\zbar_{{2}}}^{2}\right.\\
       &-1678275\,{\zbar_{{1}}}^{3}-1204665\,\zbar_{{2}}{\zbar_{{1}}}^{2}+148602\,{\zbar_{{2}}}^{2}\zbar_{{1}}+
      29375\,{\zbar_{{2}}}^{3}+2237700\,{\zbar_{{1}}}^{4}\\
       &+3455528\,\zbar_{{2}}{\zbar_{{1}}}^
      {3}+302070\,{\zbar_{{2}}}^{2}{\zbar_{{1}}}^{2}-136500\,{\zbar_{{2}}}^{3}\zbar_{{1}}-
      1678275\,{\zbar_{{1}}}^{5}-5125329\,\zbar_{{2}}{\zbar_{{1}}}^{4}\\
       &-1693290\,{\zbar_{{2}}
      }^{2}{\zbar_{{1}}}^{3}+202125\,{\zbar_{{2}}}^{3}{\zbar_{{1}}}^{2}+671310\,{\zbar_{{1}}
      }^{6}+4481781\,\zbar_{{2}}{\zbar_{{1}}}^{5}+3357810\,{\zbar_{{2}}}^{2}{\zbar_{{1}}}^{4
      }\\
       &-107721\,{\zbar_{{2}}}^{3}{\zbar_{{1}}}^{3}-111885\,{\zbar_{{1}}}^{7}-2233705\,\zbar_
      {{2}}{\zbar_{{1}}}^{6}-3969738\,{\zbar_{{2}}}^{2}{\zbar_{{1}}}^{5}-390927\,{\zbar_{{2}
        }}^{3}{\zbar_{{1}}}^{4}\\
       &+58750\,{\zbar_{{2}}}^{4}{\zbar_{{1}}}^{3}+489420\,\zbar_{{2}}{
        \zbar_{{1}}}^{7}+2634295\,{\zbar_{{2}}}^{2}{\zbar_{{1}}}^{6}+1228482\,{\zbar_{{2}}}^{3
      }{\zbar_{{1}}}^{5}-96750\,{\zbar_{{2}}}^{4}{\zbar_{{1}}}^{4}\\
       &-836700\,{\zbar_{{2}}}^{2}
      {\zbar_{{1}}}^{7}-1223340\,{\zbar_{{2}}}^{3}{\zbar_{{1}}}^{6}-62250\,{\zbar_{{2}}}^{4}
      {\zbar_{{1}}}^{5}+692430\,{\zbar_{{2}}}^{3}{\zbar_{{1}}}^{7}+122065\,{\zbar_{{2}}}^{4}
      {\zbar_{{1}}}^{6}\\
       &\left.-273015\,{\zbar_{{2}}}^{4}{\zbar_{{1}}}^{7}+29375\,{\zbar_{{2}}}^{5}{
        \zbar_{{1}}}^{6}+39750\,{\zbar_{{2}}}^{5}{\zbar_{{1}}}^{7} \right) {\Delta_{{1}}}^{-2}{\Delta
      _{{2}}}^{-2}
  \end{aligned}
\end{equation}

\begin{equation}
  \label{eq:317}
  \begin{aligned}
    f^{(3)}(z)=&-{\frac {1}{38093690880}}\, \left(
      -15917050800\,\zbar_{{1}}+456232932\,\zbar_
      {{2}}+192660441750\,{\zbar_{{1}}}^{2}\right.\\
      &+62590386030\,\zbar_{{2}}\zbar_{{1}}+
      211279484\,{\zbar_{{2}}}^{2}-1070395338600\,{\zbar_{{1}}}^{3}-794525009166\,\zbar_
      {{2}}{\zbar_{{1}}}^{2}\\
      &-114611573748\,{\zbar_{{2}}}^{2}\zbar_{{1}}-7115156792\,{\zbar_{
          {2}}}^{3}+3611036097900\,{\zbar_{{1}}}^{4}\\
      &+4485991204548\,\zbar_{{2}}{\zbar_{{1}}}
      ^{3}+1373729024769\,{\zbar_{{2}}}^{2}{\zbar_{{1}}}^{2}+172908712632\,{\zbar_{{2}}}
      ^{3}\zbar_{{1}}\\
       &+12595354536\,{\zbar_{{2}}}^{4}-8243223219000\,{\zbar_{{1}}}^{5}-
      15328771143252\,\zbar_{{2}}{\zbar_{{1}}}^{4}\\
       &-7619382247178\,{\zbar_{{2}}}^{2}{\zbar_{{
            1}}}^{3}-1534203320118\,{\zbar_{{2}}}^{3}{\zbar_{{1}}}^{2}-182097732804\,{\zbar_{{
            2}}}^{4}\zbar_{{1}}\\
       &-8683469900\,{\zbar_{{2}}}^{5}+13425941147850\,{\zbar_{{1}}}^{6
      }+35631125168634\,\zbar_{{2}}{\zbar_{{1}}}^{5}\\
       &+25991656710522\,{\zbar_{{2}}}^{2}{z
        _{{1}}}^{4}+7513251658918\,{\zbar_{{2}}}^{3}{\zbar_{{1}}}^{3}+1210003720515\,{
        \zbar_{{2}}}^{4}{\zbar_{{1}}}^{2}\\
       &+107250300570\,{\zbar_{{2}}}^{5}\zbar_{{1}}+
      2195637500\,{\zbar_{{2}}}^{6}-16018774002000\,{\zbar_{{1}}}^{7}\\
       &-59707988600022
      \,\zbar_{{2}}{\zbar_{{1}}}^{6}-61303837831056\,{\zbar_{{2}}}^{2}{\zbar_{{1}}}^{5}-
      24166432738356\,{\zbar_{{2}}}^{3}{\zbar_{{1}}}^{4}\\
      &-4928943313826\,{\zbar_{{2}}}^{4
      }{\zbar_{{1}}}^{3}-611530831590\,{\zbar_{{2}}}^{5}{\zbar_{{1}}}^{2}-26041575000\,{
        \zbar_{{2}}}^{6}\zbar_{{1}}\\
      &+14136293140200\,{\zbar_{{1}}}^{8}+74311755828120\,\zbar_{{
          2}}{\zbar_{{1}}}^{7}+106181883486822\,{\zbar_{{2}}}^{2}{\zbar_{{1}}}^{6}\\
      &+56186770195008\,{\zbar_{{2}}}^{3}{\zbar_{{1}}}^{5}+14124546987582\,{\zbar_{{2}}}^{
        4}{\zbar_{{1}}}^{4}+2183901301478\,{\zbar_{{2}}}^{5}{\zbar_{{1}}}^{3}\\
      &+141417906000
      \,{\zbar_{{2}}}^{6}{\zbar_{{1}}}^{2}-9190359208800\,{\zbar_{{1}}}^{9}-
      69493182032628\,\zbar_{{2}}{\zbar_{{1}}}^{8}\\
      &-139262199819120\,{\zbar_{{2}}}^{2}{\zbar_
        {{1}}}^{7}-99607604872014\,{\zbar_{{2}}}^{3}{\zbar_{{1}}}^{6}-31105605508380\,
      {\zbar_{{2}}}^{4}{\zbar_{{1}}}^{5}\\
      &-5666669637756\,{\zbar_{{2}}}^{5}{\zbar_{{1}}}^{4}-
      499739411500\,{\zbar_{{2}}}^{6}{\zbar_{{1}}}^{3}+4321388090250\,{\zbar_{{1}}}^{10}\\
      &
      +48656803865922\,\zbar_{{2}}{\zbar_{{1}}}^{9}+139798606371588\,{\zbar_{{2}}}^{2}{z
        _{{1}}}^{8}+137955097456758\,{\zbar_{{2}}}^{3}{\zbar_{{1}}}^{7}\\
      &+55201398291783
      \,{\zbar_{{2}}}^{4}{\zbar_{{1}}}^{6}+11744735794614\,{\zbar_{{2}}}^{5}{\zbar_{{1}}}^{5
      }+1356336265200\,{\zbar_{{2}}}^{6}{\zbar_{{1}}}^{4}\\
      &+8782550000\,{\zbar_{{2}}}^{7}{
        \zbar_{{1}}}^{3}-1414808425800\,{\zbar_{{1}}}^{11}-25014405127866\,\zbar_{{2}}{\zbar_{
          {1}}}^{10}\\
      &-106841517162632\,{\zbar_{{2}}}^{2}{\zbar_{{1}}}^{9}-149905707199956
      \,{\zbar_{{2}}}^{3}{\zbar_{{1}}}^{8}-79956636322806\,{\zbar_{{2}}}^{4}{\zbar_{{1}}}^{7
      }
  \end{aligned}
\end{equation}
\begin{equation*}
 \begin{aligned}
      &-20332329285174\,{\zbar_{{2}}}^{5}{\zbar_{{1}}}^{6}-2995433412300\,{\zbar_{{2}}}^
      {6}{\zbar_{{1}}}^{5}-77818650000\,{\zbar_{{2}}}^{7}{\zbar_{{1}}}^{4}\\
      &+300289531500
      \,{\zbar_{{1}}}^{12}+9067187221092\,\zbar_{{2}}{\zbar_{{1}}}^{11}+60845356108857\,
      {\zbar_{{2}}}^{2}{\zbar_{{1}}}^{10}\\
      &+126488366264360\,{\zbar_{{2}}}^{3}{\zbar_{{1}}}^{9
      }+93969651592314\,{\zbar_{{2}}}^{4}{\zbar_{{1}}}^{8}+29701731464910\,{\zbar_{{2}}}
      ^{5}{\zbar_{{1}}}^{7}\\
      &+5375737341495\,{\zbar_{{2}}}^{6}{\zbar_{{1}}}^{6}+
      305868024000\,{\zbar_{{2}}}^{7}{\zbar_{{1}}}^{5}-35787036600\,{\zbar_{{1}}}^{13}\\
      &-2148868232604\,\zbar_{{2}}{\zbar_{{1}}}^{12}-24743599592694\,{\zbar_{{2}}}^{2}{\zbar_{
          {1}}}^{11}-80849417068920\,{\zbar_{{2}}}^{3}{\zbar_{{1}}}^{10}\\
      &-88089834084720
      \,{\zbar_{{2}}}^{4}{\zbar_{{1}}}^{9}-36636047127000\,{\zbar_{{2}}}^{5}{\zbar_{{1}}}^{8
      }-7904357952642\,{\zbar_{{2}}}^{6}{\zbar_{{1}}}^{7}\\
      &-752243946300\,{\zbar_{{2}}}^{7
      }{\zbar_{{1}}}^{6}+1655832150\,{\zbar_{{1}}}^{14}+286405678230\,\zbar_{{2}}{\zbar_{{1}
        }}^{13}\\
      \phantom{f^{(3)}(z)=}&+6642190971806\,{\zbar_{{2}}}^{2}{\zbar_{{1}}}^{12}+37427283757680\,{\zbar_
        {{2}}}^{3}{\zbar_{{1}}}^{11}+63913185937407\,{\zbar_{{2}}}^{4}{\zbar_{{1}}}^{10}\\
      &+37575505804186\,{\zbar_{{2}}}^{5}{\zbar_{{1}}}^{9}+9831162295782\,{\zbar_{{2}}}^{6
      }{\zbar_{{1}}}^{8}+1360789452540\,{\zbar_{{2}}}^{7}{\zbar_{{1}}}^{7}\\
      &+13173825000\,
      {\zbar_{{2}}}^{8}{\zbar_{{1}}}^{6}-14575439970\,\zbar_{{2}}{\zbar_{{1}}}^{14}-
      1005306836100\,{\zbar_{{2}}}^{2}{\zbar_{{1}}}^{13}\\
      &-11572589903500\,{\zbar_{{2}}}^{
        3}{\zbar_{{1}}}^{12}-34202435930730\,{\zbar_{{2}}}^{4}{\zbar_{{1}}}^{11}-
      30897046296546\,{\zbar_{{2}}}^{5}{\zbar_{{1}}}^{10}\\
      &-10439016904684\,{\zbar_{{2}}}^
      {6}{\zbar_{{1}}}^{9}-1915988002740\,{\zbar_{{2}}}^{7}{\zbar_{{1}}}^{8}-77206500000
      \,{\zbar_{{2}}}^{8}{\zbar_{{1}}}^{7}\\
      &+56821108680\,{\zbar_{{2}}}^{2}{\zbar_{{1}}}^{14}+
      2028431619060\,{\zbar_{{2}}}^{3}{\zbar_{{1}}}^{13}+12418135213655\,{\zbar_{{2}}}^{
        4}{\zbar_{{1}}}^{12}\\
      &+19333958170350\,{\zbar_{{2}}}^{5}{\zbar_{{1}}}^{11}+
      9305421434772\,{\zbar_{{2}}}^{6}{\zbar_{{1}}}^{10}+2108401264068\,{\zbar_{{2}}}^{7
      }{\zbar_{{1}}}^{9}\\
      &+187661061000\,{\zbar_{{2}}}^{8}{\zbar_{{1}}}^{8}-129039404760\,
      {\zbar_{{2}}}^{3}{\zbar_{{1}}}^{14}-2587173466500\,{\zbar_{{2}}}^{4}{\zbar_{{1}}}^{13}\\
      &-8403448711600\,{\zbar_{{2}}}^{5}{\zbar_{{1}}}^{12}-6725788007592\,{\zbar_{{2}}}^{
        6}{\zbar_{{1}}}^{11}-1892891215014\,{\zbar_{{2}}}^{7}{\zbar_{{1}}}^{10}\\
      &-277132387700\,{\zbar_{{2}}}^{8}{\zbar_{{1}}}^{9}+188761664700\,{\zbar_{{2}}}^{4}{z
        _{{1}}}^{14}+2155595370600\,{\zbar_{{2}}}^{5}{\zbar_{{1}}}^{13}\\
      &+3522052783964
      \,{\zbar_{{2}}}^{6}{\zbar_{{1}}}^{12}+1456170527574\,{\zbar_{{2}}}^{7}{\zbar_{{1}}}^{
        11}+293531487060\,{\zbar_{{2}}}^{8}{\zbar_{{1}}}^{10}\\
      &+8782550000\,{\zbar_{{2}}}^{9
      }{\zbar_{{1}}}^{9}-185488839900\,{\zbar_{{2}}}^{5}{\zbar_{{1}}}^{14}-1165840766580
      \,{\zbar_{{2}}}^{6}{\zbar_{{1}}}^{13}\\
      &-868288332856\,{\zbar_{{2}}}^{7}{\zbar_{{1}}}^{12
      }-221357393880\,{\zbar_{{2}}}^{8}{\zbar_{{1}}}^{11}-25123350000\,{\zbar_{{2}}}^{9}
      {\zbar_{{1}}}^{10}\\
      &+123701472720\,{\zbar_{{2}}}^{6}{\zbar_{{1}}}^{14}+389322265500
      \,{\zbar_{{2}}}^{7}{\zbar_{{1}}}^{13}+124275425135\,{\zbar_{{2}}}^{8}{\zbar_{{1}}}^{12
      }\\
      &+23389674000\,{\zbar_{{2}}}^{9}{\zbar_{{1}}}^{11}-55116605880\,{\zbar_{{2}}}^{7}{
        \zbar_{{1}}}^{14}-69934264260\,{\zbar_{{2}}}^{8}{\zbar_{{1}}}^{13}\\
      &-15944383000\,{z
        _{{2}}}^{9}{\zbar_{{1}}}^{12}+15644258910\,{\zbar_{{2}}}^{8}{\zbar_{{1}}}^{14}+
      3981361650\,{\zbar_{{2}}}^{9}{\zbar_{{1}}}^{13}\\
      &+2195637500\,{\zbar_{{2}}}^{10}{\zbar_{
          {1}}}^{12}-2542777650\,{\zbar_{{2}}}^{9}{\zbar_{{1}}}^{14}+306075000\,{\zbar_{{2}}
      }^{10}{\zbar_{{1}}}^{13}\\
      &\left.+178731000\,{\zbar_{{2}}}^{10}{\zbar_{{1}}}^{14}
    \right) { \Delta_{{1}}}^{-4}{\Delta_{{2}}}^{-4}
  \end{aligned}
\end{equation*}

\section{GV invariants}
\label{sec:gv-invariants}

\begin{sidewaystable}
  \centering
   $
\begin{tiny}
   \begin {array}{|c|cccccccc|} \hline d_1 \setminus d_2  &0&1&2&3&4&5&6&7
\\[5pt] \hline0&0&540&540&540&540&540&540&540
\\[5pt] 1&3&-1080&143370&204071184&21772947555&
1076518252152&33381348217290&746807207168880\\[5pt] 2&-6&
2700&-574560&74810520&-49933059660&7772494870800&31128163315047072&
8211715737128556480\\[5pt] 3&27&-17280&5051970&-913383000
&224108858700&-42712135606368&4047949393968960&-16612333123572659520
\\[5pt] 4&-192&154440&-57879900&13593850920&-
2953943334360&603778002921828&-90433961251273800&50057390316302661600
\\[5pt] 5&1695&-1640520&751684050&-218032516800&
51350781706785&-11035406089270080&2000248139674298880&-
541531457497667187360\\[5pt] 6&-17064&19369800&-
10500261120&3630383423100&-967920854160960&224651517028866252&-
45689218327425589920&10071417619296745378920\\[5pt] 7&
188454&-245635200&153827405370&-61789428573120&18707398902511245&-
4765797079033190400&1064787653240073455400&-230224103349955979141880\\[5pt]
\hline
\end {array} 
\end{tiny}
  $
  \caption{$g=0$}
\end{sidewaystable}
\begin{sidewaystable}
  \centering
   $
\begin{tiny}
\begin {array}{|c|cccccccc|} \hline d_1 \setminus d_2&0&1&2&3&4&5&6&7
\\[5pt] \hline 0&0&3&0&0&0&0&0&0\\[5pt] 1&0&-6&2142
&-280284&-408993990&-44771454090&-2285308753398&-73398848219076
\\[5pt] 2&0&15&-8568&2126358&521854854&1122213103092&
879831736792200&205929022209626928\\[5pt] 3&-10&4764&-
1079298&152278992&-16704086880&-3328467399468&1252978673849946&-
556349234873466744\\[5pt] 4&231&-154662&48907800&-
9759419622&1591062429648&-186415241060547&8624795296820760&
2067149471538742920\\[5pt] 5&-4452&3762246&-1510850250&
385304916960&-76672173887766&12768215950604070&-1663415916220743876&
220904813068369853736\\[5pt] 6&80958&-82308270&
40028268276&-12433493287620&2931354541290318&-578520552756118977&
96321811855350031992&-15333848730658632865302\\[5pt] 7&-
1438086&1707634920&-974938365558&357248310744312&-97937943585729324&
22144022444440264176&-4288880126137360757400&762495977216972967628344\\[5pt]
\hline
\end {array}
\end{tiny}
  $
  \caption{$g=1$}
\end{sidewaystable}

\begin{sidewaystable}
  \centering
   $
\begin{tiny}
\begin {array}{|c|cccccccc|} \hline d_1 \setminus d_2&0&1&2&3&4&5&6&7
\\[5pt] \hline 0&0&0&0&0&0&0&0&0\\[5pt] 1&0&0&9&-
3192&412965&614459160&68590330119&3587118690600\\[5pt] 2&0
&0&-36&20826&-5904756&-47646003780&-80065270602672&-36393689644146360
\\[5pt] 3&0&27&-16884&4768830&-818096436&288137120463&
67873415627151&45583988161896702\\[5pt] 4&-102&57456&-
15452514&2632083714&-320511624876&18550698291252&780000198300540&-
251496603078253344\\[5pt] 5&5430&-4032288&1430896428&-
323858122812&55058565096630&-7249216518163620&691264676523200805&-
39745849558901142924\\[5pt] 6&-194022&177495894&-
77872799952&21874076033328&-4595039844606324&780316191323388252&-
108001731472892477172&12700932052931799955182\\[5pt] 7&
5784837&-6277761198&3280241914893&-1101478942766574&274831572910592142
&-55535640852991791852&9409679296993051279011&-
1395184265801287057499886\\[5pt]
\hline
\end {array}
\end{tiny}
  $
  \caption{$g=2$}
\end{sidewaystable}

\begin{sidewaystable}
  \centering
   $
\begin{tiny}
\begin {array}{|c|cccccccc|} \hline d_1 \setminus d_2&0&1&2&3&4&5&6&7
\\[5pt] \hline 0&0&0&0&0&0&0&0&0\\[5pt] 1&0&0&0&-12
&4230&-541440&-820457286&-93230754660\\[5pt] 2&0&0&0&66&-
45729&627574428&3776946955338&3154704156093636\\[5pt] 3&0
&0&-72&48036&-14756490&297044064&-7900517344212&-6362380878728364
\\[5pt] 4&15&-7236&1638918&-226431351&20419274259&-
719284158099&236091664016826&-65579326297771734\\[5pt] 5&
-3672&2417742&-764921214&154856849136&-22866882491772&2493418732350750
&-194361733345447458&11145513580945101792\\[5pt] 6&290853
&-240662448&95825798874&-24497597694651&4625681034438657&-
687273507534149145&81177932356719743208&-7626799243005256969200
\\[5pt] 7&-15363990&15291362682&-7342199188434&
2269518622807320&-518201778138767424&94360213143715120002&-
14152475137808110529514&1791485123982485159044584\\[5pt]
\hline\end {array}
\end{tiny}
  $
  \caption{$g=3$}
\end{sidewaystable}

\clearpage

\section{Modular forms}
\subsection{Definitions}
We summarize the definitions of the modular objects appearing in this work. 
\begin{equation}
\eta(\tau)=q^{\frac{1}{24}}\prod_{n=1}^\infty(1-q^n), \quad \Delta(\tau)=\eta(\tau)^{24}
\end{equation}
and transforms according to
\begin{equation}\label{etatrafo}
\eta(\tau+1)=e^{\frac{i\pi}{12}}\eta(\tau),\qquad \eta\left(-\frac{1}{\tau}\right)=\sqrt{\frac{\tau}{i}}\,\eta(\tau).
\end{equation}

The Eisenstein series are defined by
\begin{equation}\label{eisensteinseries}
E_k(\tau)=1-\frac{2k}{B_k}\sum_{n=1}^\infty\frac{n^{k-1}q^n}{1-q^n},
\end{equation}
where $B_k$ denotes the $k$-th Bernoulli number. $E_k$ is a modular form of weight $k$ for $k>2$ and even. The discriminant form is
\begin{equation}
  \label{eq:29}
  \Delta(\tau) = \frac{1}{1728}\left({E_4}(\tau)^3-{E_6}(\tau)^2\right) = \eta(\tau)^{24}.
\end{equation}

The modular completion of the holomorphic Eisenstein series $E_2$ has the form
\begin{equation}
 \widehat{E_2}(\tau) = E_2(\tau) -\frac{3}{\pi \textrm{Im}\tau} \, .
 \end{equation}
 
\subsection{Expansions of $f^{(g)}_n$}
\label{appendix:expansion}

\begin{align}
  \label{eq:ModularForms}
  f^{(0)}_1 =& \frac{1}{48}\Delta^{-\frac{3}{2}} E_{{4}} \left( 113\,{E_{{6}}}^{2}+31\,{E_{{4}}}^{3} \right) \\
  f^{(0)}_2 =& \frac{1}{221184}\Delta^{-3} \left(E_{{4}}E_{{6}} \left( 196319\,{E_{{6}}}^{4}+755906\,{E_{{6}}}^{2}{E_{{
4}}}^{3}+208991\,{E_{{4}}}^{6} \right)\right.\notag\\
& \left.+4\,{E_{{4}}}^{2} \left( 113\,{
E_{{6}}}^{2}+31\,{E_{{4}}}^{3} \right) ^{2}E_{{2}}\right)\\
  f^{(0)}_3 =& \frac{1}{557256278016}\Delta^{-\frac{9}{2}} \left(E_{{4}}\left( 360744024241\,{E_{{6}}}^{8}+4311836724416\,{E_{{
6}}}^{6}{E_{{4}}}^{3}\right.\right. \notag\\
&\left.+6966210848730\,{E_{{6}}}^{4}{E_{{4}}}^{6}+
1904214859592\,{E_{{6}}}^{2}{E_{{4}}}^{9}+49789907821\,{E_{{4}}}^{12}
 \right) \notag\\
&+8748\,{E_{{4}}}^{2}E_{{6}} \left( 113\,{E_{{6}}}^{2}+
31\,{E_{{4}}}^{3} \right)  \left( 196319\,{E_{{6}}}^{4}+755906\,{E_{{6
}}}^{2}{E_{{4}}}^{3}+208991\,{E_{{4}}}^{6} \right) E_{{2}}\notag\\
&\left.+17496\,{E_{
{4}}}^{3} \left( 113\,{E_{{6}}}^{2}+31\,{E_{{4}}}^{3} \right) ^
{3}{E_{{2}}}^{2}\right)\\
  f^{(1)}_1=& \frac{1}{576}\Delta^{-\frac{3}{2}}  E_{{4}} \left( 113\,{E_{{6}}}^{2}+31\,{E_{{4}}}^{3} \right) E_{{2}}\\
  f^{(1)}_2=& \frac{1}{31850496}\Delta^{-3} \left(-1322175\,{E_{{6}}}^{4}{E_{{4}}}^{3}-1941621\,{E_{{6}}}^{2}{E_{{4}}}^{
6}-21935\,{E_{{6}}}^{6}-197917\,{E_{{4}}}^{9}\right. \notag\\
&\left.+12\,E_{{4}}E_{{6}}
 \left( 196319\,{E_{{6}}}^{4}+755906\,{E_{{6}}}^{2}{E_{{4}}}^{3}+
208991\,{E_{{4}}}^{6} \right) E_{{2}}\right.\notag\\
&\left.+72\,{E_{{4}}}^{2} \left( 113\,{E
_{{6}}}^{2}+31\,{E_{{4}}}^{3} \right) ^{2}{E_{{2}}}^{2}\right)\\
  f^{(1)}_3=& \frac{1}{743008370688}\Delta^{-\frac{9}{2}} \left(8\,E_{{6}} \left( 87737816690\,{E_{{6}}}^{6}{E_{{4}}}^{3}+
355811791488\,{E_{{6}}}^{4}{E_{{4}}}^{6}\right. \right.\notag\\
&\left.+255154185422\,{E_{{6}}}^{2}{E
_{{4}}}^{9}+28404078217\,{E_{{4}}}^{12}+1388616631\,{E_{{6}}}^{8}
 \right) \notag\\
&+81\,E_{{4}} \left( 113\,{E_{{6}}}^{2}+31\,{E_{{4}}}^{
3} \right)  \left( 1322175\,{E_{{6}}}^{4}{E_{{4}}}^{3}+1941621\,{E_{{6
}}}^{2}{E_{{4}}}^{6}\right.\notag\\
&\left.+21935\,{E_{{6}}}^{6}+197917\,{E_{{4}}}^{9}
 \right) E_{{2}}-972\,{E_{{4}}}^{2}E_{{6}} \left( 113\,{E_{{6}}
}^{2}+31\,{E_{{4}}}^{3} \right)  \notag\\
& \left( 196319\,{E_{{6}}}^{4}+755906\,
{E_{{6}}}^{2}{E_{{4}}}^{3}+208991\,{E_{{4}}}^{6} \right) {E_{{2}}}^{2}\notag\\
&\left.-3240\,{E_{{4}}}^{3} \left( 113\,{E_{{6}}}^{2}+31\,{E_{{4}}}^{3
} \right) ^{3}{E_{{2}}}^{3} \right)\\
  f^{(2)}_1=& \frac{1}{69120}\Delta^{-\frac{3}{2}}\left({E_{{4}}}^{2} \left( 113\,{E_{{6}}}^{2}+31\,{E_{{4}}}^{3} \right) +5\,
E_{{4}} \left( 113\,{E_{{6}}}^{2}+31\,{E_{{4}}}^{3} \right) {E_{{2}}}^
{2}\right) \\
  f^{(2)}_2=& \frac{1}{1911029760}\Delta^{-3}\left( 2\,{E_{{4}}}^{2}E_{{6}} \left( 1540871\,{E_{{6}}}^{4}+7232114\,{E_{{6}
}}^{2}{E_{{4}}}^{3}+3336839\,{E_{{4}}}^{6} \right) \right. \notag\\
&+ \left( 9371817\,{
E_{{6}}}^{2}{E_{{4}}}^{6}+5997963\,{E_{{4}}}^{3}{E_{{6}}}^{4}+943457\,
{E_{{4}}}^{9}+109675\,{E_{{6}}}^{6} \right) E_{{2}}\notag\\
&-30\,E_{{4}}E_{{6}}
 \left( 196319\,{E_{{6}}}^{4}+755906\,{E_{{6}}}^{2}{E_{{4}}}^{3}+
208991\,{E_{{4}}}^{6} \right) {E_{{2}}}^{2}\notag\\
&\left.-280\,{E_{{4}}}^{2} \left( 
113\,{E_{{6}}}^{2}+31\,{E_{{4}}}^{3} \right) ^{2}{E_{{2}}}^{3}
\right)
\\
  f^{(2)}_3=& \frac{1}{4953389137920}\Delta^{-\frac{9}{2}}\left( 2\,{E_{{4}}}^{2} \left( 6841970275\,{E_{{6}}}^{8}+59257855181\,{E_{{6}
}}^{2}{E_{{4}}}^{9}\right. \right.\notag\\
& \left.+188946594537\,{E_{{6}}}^{4}{E_{{4}}}^{6}+
103842683975\,{E_{{6}}}^{6}{E_{{4}}}^{3}+1946160544\,{E_{{4}}}^{12}
 \right) \notag\\
& -36\,{E_{{4}}}^{3}E_{{6}} \left( 113\,{E_{{6}}}^{2}+31\,{E_{{
4}}}^{3} \right)  \left( 1354933\,{E_{{4}}}^{6}+2482198\,{E_{{6}}}^{2}
{E_{{4}}}^{3}+475957\,{E_{{6}}}^{4} \right) E_{{2}} \notag\\
& -9\,E_{{4}} \left( 
113\,{E_{{6}}}^{2}+31\,{E_{{4}}}^{3} \right)  \left( 954989\,{E_{{4}}}
^{9}+9455889\,{E_{{6}}}^{2}{E_{{4}}}^{6}+6151191\,{E_{{4}}}^{3}{E_{{6}
}}^{4}\right.\notag\\
& \left.+109675\,{E_{{6}}}^{6} \right) {E_{{2}}}^{2}+360\,{E_{{4}}}^{2}E
_{{6}} \left( 113\,{E_{{6}}}^{2}+31\,{E_{{4}}}^{3} \right)  \left( 
196319\,{E_{{6}}}^{4}+755906\,{E_{{6}}}^{2}{E_{{4}}}^{3}\right.\notag\\
& \left.\left.+208991\,{E_{{
4}}}^{6} \right) {E_{{2}}}^{3}+1710\,{E_{{4}}}^{3} \left( 113\,{E_{{6}
}}^{2}+31\,{E_{{4}}}^{3} \right) ^{3}{E_{{2}}}^{4}\right) \\
  f^{(3)}_1=& \frac{1}{17418240}\Delta^{-\frac{3}{2}}\left(4\,E_{{4}}E_{{6}} \left( 113\,{E_{{6}}}^{2}+31\,{E_{{4}}}^{3} \right) 
+21\,{E_{{4}}}^{2} \left( 113\,{E_{{6}}}^{2}+31\,{E_{{4}}}^{3}
 \right) E_{{2}}\right.\notag\\
& \left.+35\,E_{{4}} \left( 113\,{E_{{6}}}^{2}+31\,{E_{{4}}}^{
3} \right) {E_{{2}}}^{3}\right) \\
  f^{(3)}_2 =& \frac{1}{321052999680}\Delta^{-3}\left(E_{{4}} \left( 14470511\,{E_{{6}}}^{6}+299836579\,{E_{{4}}}^{3}{E_{{6}
}}^{4}+378756589\,{E_{{6}}}^{2}{E_{{4}}}^{6}\right.\right.\notag\\
& \left.+31120385\,{E_{{4}}}^{9}
 \right) +12\,{E_{{4}}}^{2}E_{{6}} \left( 3459163\,{E_{{6}}}^{4}+
16800202\,{E_{{6}}}^{2}{E_{{4}}}^{3}+7775707\,{E_{{4}}}^{6} \right) E_
{{2}}\notag\\
&+ \left( 767725\,{E_{{6}}}^{6}+5958407\,{E_{{4}}}^{9}+33404973\,{
E_{{4}}}^{3}{E_{{6}}}^{4}+60894687\,{E_{{6}}}^{2}{E_{{4}}}^{6}
 \right) {E_{{2}}}^{2}\notag\\
&-140\,E_{{4}}E_{{6}} \left( 196319\,{E_{{6}}}^{4
}+755906\,{E_{{6}}}^{2}{E_{{4}}}^{3}+208991\,{E_{{4}}}^{6} \right) {E_
{{2}}}^{3}\notag\\
&\left.-2100\,{E_{{4}}}^{2} \left( 113\,{E_{{6}}}^{2}+31\,{E_{{4}}}
^{3} \right) ^{2}{E_{{2}}}^{4}\right)\\
   f^{(3)}_3 =& \frac{1}{624127031377920}\Delta^{-\frac{9}{2}}\left( 2\,E_{{4}}E_{{6}} \left( 42089002745\,{E_{{6}}}^{8}+856373539390\,{E_{
{6}}}^{6}{E_{{4}}}^{3}\right.\right.\notag\\
& \left.+2773682486544\,{E_{{6}}}^{4}{E_{{4}}}^{6}+
2005074999106\,{E_{{6}}}^{2}{E_{{4}}}^{9}+260719698551\,{E_{{4}}}^{12}
 \right) \notag\\
& -27\,{E_{{4}}}^{2} \left( 113\,{E_{{6}}}^{2}+31\,{E_{{4}}}^{3
} \right)  \left( 8126451\,{E_{{4}}}^{9}+97251020\,{E_{{6}}}^{2}{E_{{4
}}}^{6}\right. \notag\\
& \left.+74249327\,{E_{{4}}}^{3}{E_{{6}}}^{4}+2870738\,{E_{{6}}}^{6}
 \right) E_{{2}}-54\,{E_{{4}}}^{3}E_{{6}} \left( 113\,{E_{{6}}}^{2}+31
\,{E_{{4}}}^{3} \right)  \left( 9472999\,{E_{{4}}}^{6}\right. \notag\\
& \left.+17291314\,{E_{{
6}}}^{2}{E_{{4}}}^{3}+3178471\,{E_{{6}}}^{4} \right) {E_{{2}}}^{2}-315
\,E_{{4}} \left( 113\,{E_{{6}}}^{2}+31\,{E_{{4}}}^{3} \right)  \left( 
180619\,{E_{{4}}}^{9}\right. \notag\\
& \left.+1815513\,{E_{{6}}}^{2}{E_{{4}}}^{6}+1092333\,{E_
{{4}}}^{3}{E_{{6}}}^{4}+21935\,{E_{{6}}}^{6} \right)
{E_{{2}}}^{3}\notag\\
& +
1890\,{E_{{4}}}^{2}E_{{6}} \left( 113\,{E_{{6}}}^{2}+31\,{E_{{4}}}^{3}
 \right)  \left( 196319\,{E_{{6}}}^{4}+755906\,{E_{{6}}}^{2}{E_{{4}}}^
{3}+208991\,{E_{{4}}}^{6} \right) {E_{{2}}}^{4}\notag\\
&\left.+12285\,{E_{{4}}}^{3}
 \left( 113\,{E_{{6}}}^{2}+31\,{E_{{4}}}^{3} \right) ^{3}{E_{{2}}}^{5}
 \right)
\end{align}

\providecommand{\href}[2]{#2}\begingroup\raggedright\endgroup

\end{document}